\documentclass[%
 reprint,
 amsmath,amssymb,
 aps,
 pre,
]{revtex4-1}
\usepackage[font=small,format=plain,labelfont=bf,up,textfont=normal,up,justification=justified,singlelinecheck=false]{caption}
\usepackage{subcaption}
\usepackage{verbatim}
\usepackage{color}
\usepackage{ragged2e}
\usepackage{graphicx}
\usepackage{dcolumn}
\usepackage{bm}
\usepackage{amsmath}
\usepackage{hyperref}
\usepackage[mathlines]{lineno}

\begin{document}

\preprint{APS/123-QED}

\title{Localization and instability in sheared granular materials: Role of friction and vibration}

\author{Konik R. Kothari}
 \email{kkothar3@illinois.edu}
\affiliation{%
	Mechanical Engineering Department\\
 University of Illinois at Urbanna-Champaign
}%

\author{Ahmed E. Elbanna}
 \homepage{https://publish.illinois.edu/mcslabuiuc/}
 \email{elbanna2@illinois.edu}
\affiliation{
 Civil and Environmental Engineering Department\\
 University of Illinois at Urbana - Champaign\
}%

\date{\today}

\begin{abstract}
Shear banding and stick-slip instabilities have been long observed in sheared granular materials. Yet, their microscopic underpinnings, interdependencies and variability under different loading conditions have not been fully explored. Here, we use a non-equilibrium thermodynamics model, the Shear Transformation Zone theory, to investigate the dynamics of strain localization and its connection to stability of sliding in sheared, dry, granular materials. We consider frictional and frictionless grains as well as presence and absence of acoustic vibrations. Our results suggest that at low and intermediate strain rates, persistent shear bands develop only in the absence of vibrations. Vibrations tend to fluidize the granular network and de-localize slip at these rates. Stick-slip is only observed for frictional grains and it is confined to the shear band. At high strain rates, stick-slip disappears and the different systems exhibit similar stress-slip response. Changing the vibration intensity, duration or time of application alters the system response and may cause long-lasting rheological changes. We analyse these observations in terms of possible transitions between rate strengthening and rate weakening response facilitated by a competition between shear induced dilation and vibration induced compaction. We discuss the implications of our results on dynamic triggering, quiescence and strength evolution in gouge filled fault zones.

\end{abstract}

\maketitle


\section{Introduction}
Understanding the dynamics of stick-slip instabilities in sheared granular materials is a fundamental and long-standing challenge in earthquake source physics. In the case of mature faults that have accumulated hundreds of meters of slip throughout their active history, there is little doubt that shear deformation localizes dynamically to very thin zones \citep{chester1993internal,chester1998ultracataclasite,scholz2002mechanics,ramsey2004hybrid,noda2005thermal,morrow2000effect,
ben2003characterization,segall2006does}. Unravelling how slip localizes in such thin zones within a wider region of damaged materials and how the thickness of this shear band evolves as a function of different loading conditions and what the implications for the stability of sliding are, may hold the keys for understanding a number of outstanding problems in geophysics such as origins of earthquake complexity and the nature of heat flow paradox.

In recent years, there has been a surge of interest in understanding the role of acoustic vibrations in controlling granular rheology. This is in part due to increased observations of dynamic earthquake triggering\citep{brodsky2014uses,gomberg2001earthquake,gomberg2005seismology}. Laboratory-scale experiments and seismological observations suggest that a key role in dynamic earthquake triggering may be played by the non-linear dynamics of fault gouge accumulating at the core of fault zones\citep{johnson2008effects,johnson2005nonlinear}. Furthermore, insight from discrete element modelling\citep{ferdowsi2015acoustically} and small-scale direct shear experiments suggest that acoustic vibrations may lead to long lasting changes in granular rheology which in turn may change seismic observables such as stress drop, inter-event times and earthquake magnitudes in the period following the cessation of slip compared to the un-vibrated cases\citep{ferdowsi2015acoustically}. 

In this paper, we numerically examine the stability of sliding of a layer of granular materials in the absence and presence of acoustic vibrations. The layer fabric may be chosen to reflect frictionless or frictional particles. Our primary theoretical tool is a recent formulation of the Shear Transformation Zone (STZ) theory by Lieou \textit{et al}\citep{lieou2014grain,lieou2014shear,lieou2015stick} that explicitly incorporates the effect of inter-particle friction as well as internal and external vibration sources on granular rearrangements.  We extend this formulation to account for the potential of strain localization and athermal shear band development. This will enable exploring some of the implications of microscopic processes as well as acoustic vibrations on the transition between localized slip and distributed deformation as well as the switch between unstable and stable slip.

The shear transformation zone (STZ) theory is a continuum model of plastic deformation in amorphous solids that quantifies local configurational disorder \citep{falk1998dynamics}. The basic assumption in the theory is that plastic deformation occurs at rare non-interacting localized spots known as shear transformation zones (STZs). An internal state variable, the effective temperature, describes fluctuations in the conﬁgurational states of the granular material (i.e., a measure of local entropy) and controls the density of STZs \citep{langer2004dynamics,haxton2007activated,manning2009rate,bouchbinder2009nonequilibrium}. Effective temperature can be related to the system porosity \citep{lieou2012nonequilibrium,lieou2014grain,lieou2014shear,lieou2015stick}. This approach coarse-grains granular simulations while retaining important physical concepts.

The volume of a granular layer, under the combined effects of shear and vibrations, evolves under the competing influences of shear induced dilation and vibration induced compaction. As a result, the granular layer volume may exhibit a non-monotonic dependence on strain rate \citep{elst2012auto,lieou2014grain,lieou2014shear,lieou2015stick}. While shear induced dilation has been coined as rate strengthening, the non-montonic volume response in the presence of vibration points to possible transition into rate weakening and accordingly loss of sliding stability. This has been successfully demonstrated by Lieou \textit{et al.} \citep{lieou2016dynamic,lieou2015stick} who explored stick-slip dynamics under a wide range of strain rates and confining pressures and discovered a transition from fast stick-slip to slow stick-slip and ultimately stable sliding as a function of increasing vibration intensity. However, in these previous studies it was observed that the stability properties before the application of external vibrations are recovered upon cessation of vibration and apparently no lasting rheological changes occurred. This points to the importance of including additional physics related to the evolution of gouge internal state that may have different relaxation time scale than the scale of granular rearrangements. Two possible candidates are shear bands and non-linear changes in gouge elastic properties as a function of porosity.

The new contribution in this this paper is the inclusion of strain localization potential in theoretical models of fault gouge subjected to vibration and shear. This has important geophysical implications for earthquake nucleation and propagation, as well as assessing fault zone susceptibility for dynamic triggering and understanding seismic patterns. The model is also an important step towards integrating effects of vibrations in multi-scale models of fault zone inelasticity and dynamic earthquake ruptures.

The remainder of the paper is organized as follows. In section 2 we review the basic elements of the STZ theory. In section 3 we discuss the set-up for the one dimensional fault zone model, summarize the selected parameters and provide an overview for the numerical procedure. In section 4, we investigate the predictions of the model for the emergence of stick-slip instabilities and the evolution of shear bands under different strain rates and vibration intensities. In section 5, we discuss the implications of this model for earthquake dynamics and seismic patterns and summarize our conclusions.

\section{Review of STZ theory} \label{review}

Granular particles can move and rearrange in response to applied stress. Molecular dynamics simulations of such amorphous materials show that plastic irreversible deformation is localized in regions called shear transformation zones (STZs) \citep{falk1998dynamics}. These regions undergo irreversible conﬁgurational rearrangement in a sporadic, transient and spatially isolated fashion. These re-arrangements occur in persistently noisy environments due to mechanical (and thermal) fluctuations bringing groups of grains in and out of local clusters. STZs are assumed to be dynamic and anisotropic entities whose density and orientation is governed by the loading history.

\begin{figure} \label{particlepicture}
\includegraphics[scale=0.25]{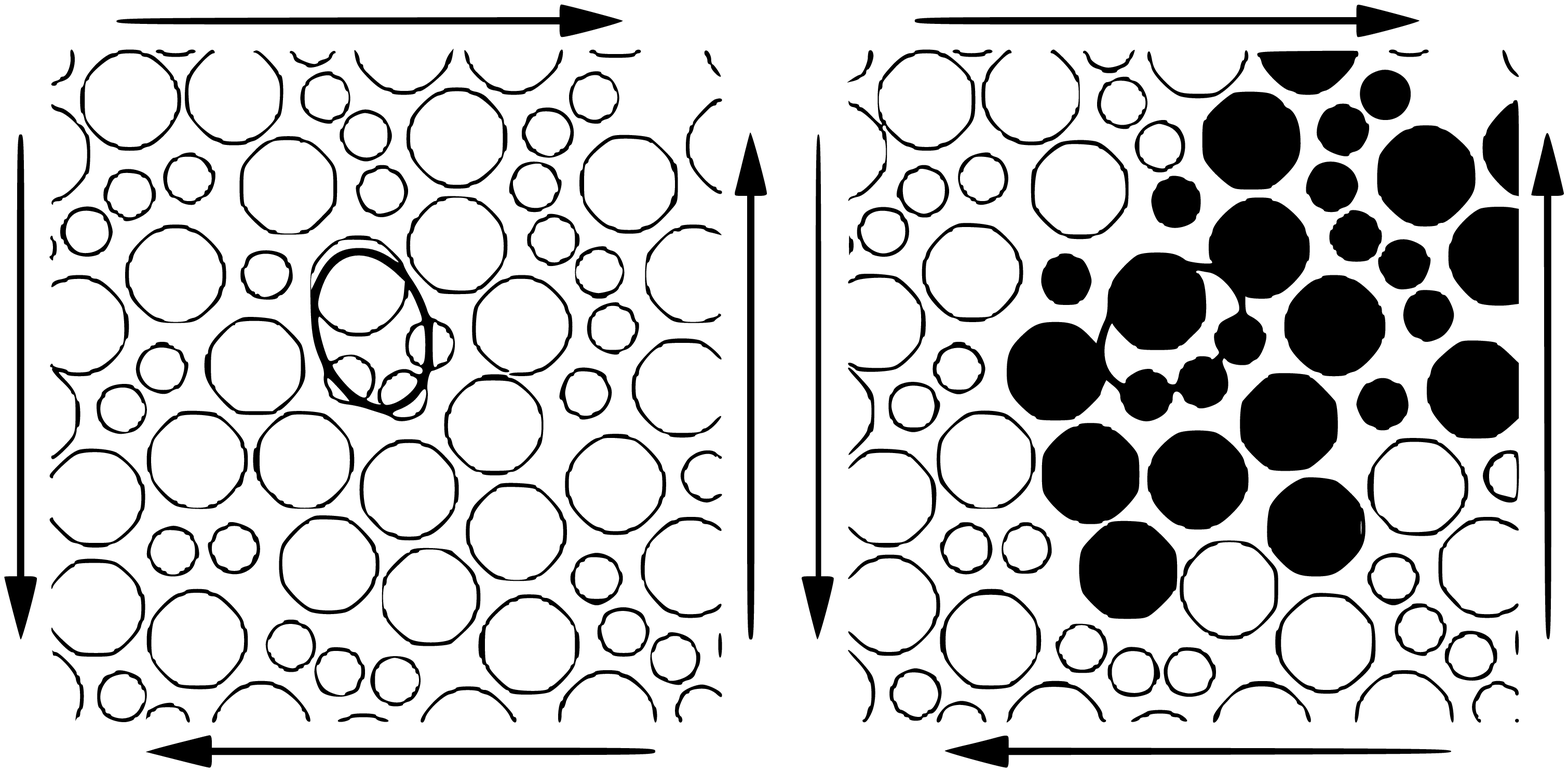}
\caption{Schematic of an STZ transition. Under applied shear ( the shear sense is given by the arrows) a cluster of particles (identified by the oval region on the left) rearrange by changing their local topology (on the right). Picture from \citep{falk1998dynamics}}
\end{figure}

If a shear transformation occurs, the STZ resist further shear strain in the shearing direction. A shear transformation represents flipping between two orientations, anti-aligned $(-)$ and aligned $(+)$ to the direction of principal shear stresses \citep{falk1998dynamics}. In an STZ, particles exchange nearest neighbor relationships. This event generates a quantum of local plastic strain. The macroscopic plastic strain is the cumulative result of many such local events. Once flipped, STZs cannot further deform in the same direction. Instead, they are continuously created and destroyed in order to further accommodate plastic strain within the material. The amount of conﬁgurational disorder in the system is characterized by a single state variable: the effective temperature, $\chi$. The effective temperature is formally defined as the change in the system potential energy (or volume) per unit change in the system entropy \citep{bouchbinder2009nonequilibrium,lieou2012nonequilibrium}.
\begin{equation} \label{unnorm_chidef}
X = \left( \dfrac{\partial V}{\partial S_c} \right)_{\zeta_i}
\end{equation}
We normalize this $X$ by the excess volume per STZ, $v_z$ (i.e. $\chi = \dfrac{X}{v_z}$). This normalized quantity can be related to the granular layer's porosity and is referred to as compactivity in the later parts of this paper. A fundamental result in the STZ theory is that the continuous creation and annihilation of the STZs drive their density $\Lambda$ toward a Boltzmann distribution $\Lambda \ \propto \ exp(-1/\chi)$ \citep{langer2008shear,manning2009rate,lieou2012nonequilibrium}. In this work, we introduce a non-dimensional measure of shear stress, $s$, $\mu = \dfrac{s}{p}$ where $p$ is the confining pressure. We use $a$ to denote the characteristic grain size of the grains and $\rho_G$ to denote its material density. 
The master rate equations of STZs can be written as-  \citep{falk1998dynamics,lieou2012nonequilibrium}:
\begin{subequations} \label{basiceqn}
\begin{equation} \label{masterrate}
\tau \dot{N}_{\pm} = R(\pm\mu,\chi)N_{\mp} - R(\mp\mu,\chi)N_{\pm} + \tilde{\Gamma} \left( \dfrac{1}{2} N^{eq} - N_{\pm} \right)
\end{equation}
\begin{equation}\label{initplastic}
\tau \dot{\gamma}^{pl} = \dfrac{2\epsilon_0}{N}\left(R(\mu,\chi)N_- - R(-\mu,\chi)N_+\right)
\end{equation}
\end{subequations}
where $\tau = a\sqrt{\frac{\rho_G}{p}}$ is the inertial time scale and $\tilde{\Gamma} = \Gamma + \rho$ is the attempt frequency. Here, $\Gamma$ and $\rho$ denote the mechanical and vibration noise strengths respectively. We do not include friction in the attempt frequency as it does not explicitly open or close voids. $N_{\pm}$ indicates STZs oriented in the $+$ (aligned with shear direction) or $-$ (anti-aligned with shear direction) direction respectively, $R$ denotes their flipping rates and $N_{eq}$ indicates the equilibrium number of STZs. The first two terms of eqn. \ref{masterrate} represent the entropy-conserving process of flipping of STZs from one state to another. The third term represents the creation and annhiliation of STZs. It must be noted that a higher attempt frequency (or stronger mechanical and vibrational noise) leads to a faster creation or annhilation of voids for the same departure of '+' or '-' oriented STZ population from its equilibrium value. In a system deforming plastically under shear, neither of $N_{+}$ or $N_{-}$  will be equal to $N_{eq}/2$ but at steady state $N_{+}+N_{-}=N_{eq}$. Further, note that it is assumed that both orientations of STZs have an equal likelihood of creation and annhiliation as $\tilde{\Gamma}$ does not depend on orientation of STZs. 

To get \ref{initplastic} from eqn. \ref{masterrate}, one must note that all plastic strain is accomodated through shear transformations or flipping of STZs.  It is assumed that the annihilation or creation of STZs do not generate plastic strain rate. $\epsilon_0$ is an average shear strain increment of order unity and $\tau_0$ is the time scale for particle rearrangements. Eqn. \ref{initplastic} then, represents a weighted sum over all flip events. For a thorough treatment, please refer \citep{langer2008shear}.

Following, previous discussions, we introduce the following intensive quantities:

\begin{subequations} \label{micro}
\begin{equation}
    \Lambda = \dfrac{N_+ + N_-}{N} ;\ m = \dfrac{N_+ - N_-}{N_+ + N_-}
\end{equation}
\begin{equation}
    C = \dfrac{1}{2}\left(R_+ + R_-\right) ;\ \mathcal{T} = \dfrac{R_+ - R_-}{R_+ + R_-}
\end{equation}
\end{subequations}

where $\Lambda$ is the STZ density, $m$ the orientational bias, $R_{\pm} = R(\pm\mu,\chi)$ is the rate of STZs flipping {\bf to} '+' or '-' state and $N$ is the total number of grains/particles in the same volume. The rate parameter $C$ describes the flipping rate bias and $\mathcal{T}$ is the orientational rate bias.

Using eqns. \ref{micro}, the eqns. \ref{basiceqn} can be written in terms of intensive quantities:
\begin{subequations} \label{micro_all_rate}
\begin{equation} \label{rate_pls}
    \tau \dot{\gamma}_{pl} = 2\epsilon_0 \Lambda C(\mu,\chi) [\mathcal{T}(\mu,\chi) - m]
\end{equation}
\begin{equation} \label{Lambda}
    \tau \dot{\Lambda} = \tilde{\Gamma} ( \Lambda_{eq} - \Lambda)
\end{equation}
\begin{equation} \label{meq_first}
    \tau \dot{m} = 2C(\mu,\chi)[\mathcal{T}(\mu,\chi) - m] - \tilde{\Gamma}m - \tau \dfrac{\dot{\Lambda}}{\Lambda}m
\end{equation}
\end{subequations}
where $\epsilon_0$ is a constant.

The expression for the mechanical noise strength term, $\Gamma$ can be derived by invoking the Pechenik hypothesis \citep{langer2003dynamics} that states that $\Gamma$ is directly proportional to the plastic work done per STZ:
\begin{equation} 
	\Gamma = \dfrac{\tau\mu\dot{\gamma}_{pl}}{\epsilon_0\mu_0\Lambda_{eq}} = \dfrac{2\mu}{\mu_0}R_0\left(\mathcal{T}(\mu,\chi) - m\right)
	\label{mechnoise}
\end{equation} 
where we replaced $C(\mu,\chi)$ by some constant $R_0$ as when $\mu$ is less than unity, the STZ flipping rate should not depend strongly on shear stress as was suggested in prior investigations \citep{lieou2014shear, lieou2015stick} (see \hyperref[AppendixA]{Appendix} for more details). 

Now, let $W$ denote the inelastic mechanical work done on the system and $q = \tau\dot{\gamma}$ the non-dimensionalized strain rate. Then, in the absence of vibration or internal friction, $\chi$ evolves according to the following equation \citep{lieou2012nonequilibrium, lieou2014grain, lieou2014shear, lieou2015stick, lieou2016dynamic}
\begin{equation} \label{chidot_ysnfnv}
    \dot{\chi} = \dfrac{W}{\tau} \left( 1 - \dfrac{\chi}{\hat{\chi}(q)}\right)
\end{equation}

This equation states that under the effect of external work, the compactivity increases monotonically to a strain rate dependent steady-state value given by $\hat{\chi}(q)$. Earlier studies\citep{langer2008shear} suggest that $\hat{\chi}(q)$ is a monotonically increasing function of the strain rate with the property that $\hat{\chi}(q) \rightarrow \chi_0$  if $q \rightarrow 0$ and $\hat{\chi}(q) \rightarrow \infty$  as $q \rightarrow q^*$ . Here $q^*$ is a characteristic inertia number at which the behavior of the granular layer transition from the dense regime to the fluid like regime. 

Intuitively, if all the particles move without changing their local topology, then this motion is reversible. In that case, the deformation of the bulk remains elastic and compactivity remains constant since $W=0$ in this limit. Irreversible non-affine transformation is characterized by arrangements that displace particles relative to one another leading to changes in their coordination number or nearest neighbor relationships. Under shear, for a particle to move past another one it has also to overcome the confinement offered by the surrounding bulk. Local dilation increases the free volume of the particle and facilitate this shear induced rearrangement. This is captured in the monotonic increase of the compactivity with continuous shearing as encapsulated in eqn \ref{chidot_ysnfnv}.

If the granular medium is vibrated but not sheared, the steady-state compactivity is determined by the vibration noise strength, $\rho$ and the compactivity evolution is driven by the vibration work rate.\citep{lieou2012nonequilibrium, lieou2014grain, lieou2014shear, lieou2015stick} 
\begin{equation} \label{chidot_nsnfyv}
    \dot{\chi} = \dfrac{Y}{\tau} \left( 1 - \dfrac{\chi}{\tilde{\chi}(\rho)}\right)
\end{equation}
 
where $\tilde{\chi}(\rho)$ the steady-state value of $\chi$ at a given vibration noise strength. According to past studies\citep{daniels2005hysteresis,daniels2006characterization}, the quantity $\rho$ is proportional to the vibrational amplitude and the square of the frequency. Furthermore, experiments\citep{nowak1998density,knight1995density} suggest that  $\tilde{\chi}(\rho_1) < \tilde{\chi}(\rho_2)$ whenever $\rho_1 > \rho_2$  signifying that higher vibration intensities lead to increased compaction (lower compactivity).  

Intuitively, vibrations add random fluctuations to the particles motion which should disrupt particles coordination and open up inter-particle spacing. However, under the effect of gravity or under constant pressure loading,  particles must possess a minimum coordination number to transfer the applied loading. To facilitate this, particles move around to fill up these spaces leading again to local topological changes, density increases and overall compaction.  Earlier experiments \citep{jaeger1992physics,elst2012auto} provided evidence for such a picture. Mathematically, this is captured by eqn. \ref{chidot_nsnfyv} in that whenever $Y \neq 0$ and $\chi>\chi(\rho)$, the compactivity decreases.

If the granular material is vibrated and sheared, one may then write a general equation for the evolution of the compactivity as follows:
\begin{equation} \label{chidot_ysnfyv}
    \dot{\chi} = \dfrac{W}{\tau} \left( 1 - \dfrac{\chi}{\hat{\chi}(q)}\right) + 
        \dfrac{Y}{\tau} \left( 1 - r(q, \rho)\dfrac{\chi}{\tilde{\chi}(\rho)}\right)
\end{equation}

where $Y$ is the work rate done by the external vibrations and  $r$ is an interpolation function with the property that: $r(0,\rho)=1$ and $r(\infty,\rho)=0$. The first condition ensures that steady-state $\chi^{ss}= \tilde{\chi}(\rho)$ in the absence of shear, while the second condition ensures that in the limit of high strain rates the steady-state compactivity $\chi^{ss}= \hat{\chi}(q)$, consistent with experimental observations \citep{elst2012auto}. This may be made more explicit by setting $ \dot{\chi} $ to zero and solving for the steady-state compactivity in the above equation:
\begin{equation} \label{sschi}
    \dfrac{1}{\chi} = \dfrac{A}{\hat{\chi}(q)} + \dfrac{B}{\tilde{\chi}(\rho)}
\end{equation}
where $A = \frac{W}{W+Y}$ and $B = \frac{rY}{W+Y}$.  In a previous study\citep{lieou2015stick}, it was shown that $r = \text{exp}(-\frac{q^2}{\rho})$ is a plausible interpolation function.
Experiments\citep{elst2012auto} suggest that angular grains upon shearing generate internal acoustic vibrations that auto compact the granular layer. To model this effect, Lieou \textit{et al.}\citep{lieou2015stick} accounted for inelastic collisions between angular grains through the introduction of a frictional dissipation term in the thermodynamics set-up. The rationale is that the inter-particle friction should be reducing the disorder state of the system because it restricts the configurational space for the particles. The evolution equation of compactivity is modified to:

\begin{equation} \label{chidot_ysyfyv}
    \dot{\chi} = \dfrac{W}{\tau} \left( 1 - \dfrac{\chi}{\hat{\chi}(q)}\right) + 
        \dfrac{Y}{\tau} \left( 1 - r(q, \rho)\dfrac{\chi}{\hat{\chi}(\rho)}\right) 
        -\dfrac{F}{\tau} \dfrac{\chi}{\hat{\chi}(q)}
\end{equation}

Here $F$  is the frictional energy dissipated per unit time through inelastic collisions. At low strain rates, this energy should scale with the contact area and square of plastic flow rate. However, at high strain rates the contact area would be reduced due to shear induced dilation and hence, a limit may be hypothesized. Therefore, we can write  $F = \xi\epsilon_0\mu_0\Lambda$, where $\xi$ varies as $\dot{\gamma}_{pl}^2$ in the limit of small strain rates and saturates to a constant value in the limit of high strain rates. We assume $\xi = \xi_0 tanh((\tau_{fr} \dot{\gamma}_{pl})^2)$. With this expression for $\xi$, Eqn. \ref{chidot_ysyfyv} predicts that that even in the absence of external acoustic vibrations, frictional interactions between the grains has a compaction effect (i.e. reducing the effective temperature) that competes with the shear induced dilation (i.e. tendency to increase effective temperature). 
If $\chi$ is spatially heterogeneous, there will be a transfer of disorder from points of high disorder to points of lower disorder. To account for this disorder diffusion, inspired by models of gradient plasticity as well as diffusion of Gibbs temperature between systems not in thermal equilibrium, the compactivity equation is modified as follows:

\begin{equation} \label{chidot_final}
\begin{split}
        \dot{\chi} = \dfrac{W}{\tau} \left( 1 - \dfrac{\chi}{\hat{\chi}(q)}\right) &+ 
        \dfrac{Y}{\tau} \left( 1 - r(q, \rho)\dfrac{\chi}{\hat{\chi}(\rho)}\right) \\
        &-\dfrac{F}{\tau} \dfrac{\chi}{\hat{\chi}(q)} + 
        \dfrac{\partial}{\partial y} \left( D\dot{\gamma}^2_{pl}\dfrac{\partial \chi}{\partial y}\right)
\end{split}
\end{equation}

Here, $D$ denotes the spatial scale of compactivity diffusion, and it scales with the square of particle size. Stability analysis (in the absence of friction and vibration)\citep{manning2007strain} shows that the feedback between the strain rate and compactivity may amplify spatial heterogeneities in the compactivity and ultimately lead to shear banding. Homogeneous deformation corresponds to $\dfrac{\partial \chi}{\partial y} = 0$. For closure of the system of equations, we assume an additive decomposition of strain rate into elastic and plastic parts. The stress rate is proportional to the rate of the elastic strain. Mathematically,

\begin{equation}
    \dot{s} = \int_{-L/2}^{L/2} \left[G \left( \dot{\gamma} - \dot{\gamma}_{pl} \right)\right]dy
\end{equation}
where, $G$ is the bulk shear modulus

\section{Model Set-up} \label{modelsetup}

We are simulating an infinite layer of dense granular material of thickness, $L$ under a uniform hydrostatic pressure, $p$ and boundary shear rate corresponding to net sliding velocity, $v$. We assume homogeneity along the horizontal dimensions, $x$ and $z$ and thus all our variables are function of the thickness ($y$-coordinate) only.
We further assume that the STZ density and the orientational bias evolve on fast time scales relative to the compactivity. This may be justified by the looking at Eqns. \ref{rate_pls} where evolution of $\dot{\gamma}_{pl}$ is directly proportional to STZ density, $\Lambda$ while that of the $m$ and $\Lambda$ is not. As a result, we can assume $m$ and $\Lambda$ to be at their instantaneous equilibrium values:

\begin{subequations}
\begin{equation} \label{lambda_eq}
    \Lambda_{eq} = 2e^{-1/\chi}
\end{equation}
\begin{equation} \label{meq}
\begin{split}
     m_{eq} = &\dfrac{\mu_0}{2\mu} \left[1 + \dfrac{\mu}{\mu_0}\mathcal{T}(\mu,\chi) + \dfrac{\rho}{2R_0}\right]\\
     & - \dfrac{\mu_0}{2}\sqrt{\left[1 + \dfrac{\mu}{\mu_0}\mathcal{T}(\mu,\chi) + \dfrac{\rho}{2R_0}\right]^2 - \dfrac{4\mu}{\mu_0}\mathcal{T}(\mu,\chi)}
\end{split}
\end{equation}
\end{subequations}
Here, $\mu_0$ is the apparent yield parameter. As the name suggests, this is the non-dimensional shear stress value that when reached in the absence of friction and vibrations, would cause yielding and initiate plastic flow. Eqn \ref{meq} may be simplified in case $\rho=0$ as:
\begin{equation} \label{simplified_meq}
    m_{eq} = \begin{cases}
             \mathcal{T}(\mu,\chi), &\text{if }  (\mu/\mu_0)\mathcal{T}(\mu,\chi) < 1 \\
             \mu_0/\mu, &\text{if }  (\mu/\mu_0)\mathcal{T}(\mu,\chi) \ge 1
            \end{cases}
\end{equation}
It follows from Eqn. \ref{simplified_meq} and Eqn. \ref{rate_pls} that $\dot{\gamma}_{pl} = 0$, if $\rho = 0$ and $(\mu/\mu_0)\mathcal{T}(\mu,\chi) < 1$, where $\mathcal{T} = tanh(\dfrac{\epsilon_0 \mu}{\epsilon_z \chi})$ \citep{lieou2015stick} (see \hyperref[AppendixA]{Appendix} for more details). Here, $\epsilon_z$ is the normalized excess volume per STZ.

In this paper, we assume that the apparent yield stress is a function of compactivity. Intuitively, as the material becomes more compacted (lower compactivity) it should be more difficult to introduce non-affine rearrangements in the material that promote yielding and vice-versa. Hence, yield stress should decrease with increasing $\chi$.  Lieou and Langer\citep{lieou2012nonequilibrium} suggested, guided by MD simulation results from \citep{haxton2011universal}, that the minimum shear stress should decrease as a function of Gibb’s temperatures to a minimum value and then remain constant. By analogy, we assume that the minimum yield stress assumes the following functional dependence on the compactivity:
\begin{equation} \label{s0}
\mu_0 = \begin{cases} A - B\chi & \chi < \chi_0 \\
        constant & \chi \ge \chi_0
        \end{cases}
\end{equation}
where we chose $\chi_0 = \hat{\chi}(q)$

We assume a spatially heterogeneous initial distribution of compactivity.
\begin{equation}
    \chi = 0.18 \left(1 + sech\left(\dfrac{ky}{L}\right)\right)
\label{chi_init}
\end{equation}
where we chose $k=10$ and $L$ describes the thickness of granular layer, which is chosen to be unity in our case. We also assume no flux conditions for the compactivity at the boundaries, that is
\begin{equation} \label{bound_cond}
\dfrac {\partial \chi}{\partial y} = 0 \bigg|_{y =\pm L/2}
\end{equation}

To summarize, we write all governing equations as below:
\begin{subequations}
\begin{equation} \label{sdot}
    \dot{s} = \int_{-L/2}^{L/2} \left[G \left( \dot{\gamma} - \dot{\gamma}_{pl} \right)\right]dy
\end{equation}
\begin{equation} \label{gammadot}
    \tau \dot{\gamma}_{pl} = 2\epsilon_{0} \Lambda R_0[\mathcal{T}(\mu,\chi) -m]
\end{equation}
\begin{equation} \label{chidot}
\begin{split}
    \tau \dot{\chi} =& \dfrac{2\epsilon_0\mu_0 e^{-1/\chi}}{\epsilon_1} \left[ \Gamma\left(1-\dfrac{\chi}{\hat{\chi}(q)}\right) - \xi\dfrac{\chi}{\hat{\chi}(q)} \right] \\ &+ \dfrac{A_0\rho}{\epsilon_1}\left[1-e^\frac{-q^2}{\rho}\dfrac{\chi}{\tilde{\chi}(\rho)}\right] \\ &+ D_0 a^2 \dot{\gamma}_{pl}^2 \dfrac{\partial^2 \chi}{\partial y^2}
\end{split}
\end{equation}
\begin{equation}
\Lambda_{eq} = 2e^{\frac{-1}{\chi}}
\end{equation}
\begin{equation}
\begin{split}
     m_{eq} = &\dfrac{\mu_0}{2\mu} \left[1 + \dfrac{\mu}{\mu_0}\mathcal{T}(\mu,\chi) + \dfrac{\rho}{2R_0}\right]\\
     & - \dfrac{\mu_0}{2\mu}\sqrt{\left[1 + \dfrac{\mu}{\mu_0}\mathcal{T}(\mu,\chi) + \dfrac{\rho}{2R_0}\right]^2 - \dfrac{4\mu}{\mu_0}\mathcal{T}(\mu,\chi)}
\end{split}
\end{equation}
\label{eqns:all}
\end{subequations}
where, $A_0$ is a vibration constant and $D_0$ is the compactivity diffusion coefficient.

We assume constant $\hat{\chi}(q) = \hat{\chi}_0 \ \forall \ q$ as the strain rates considered here correspond to a dimensionless flow rate, $q$ much lower than the critical flow rate of $q \approx 1$ beyond which the granular layer enters the fluid-like regime. In particular, $q \approx 1$  corresponds to a plastic strain rate of $\dot{\gamma} = \frac{q}{\tau} \approx 10^6-10^7 /s$ (for the parameters in this work)  which is much higher than the strain rates considered here. Hence, a constant $\hat{\chi}(q)$ is reasonable.

\begin{table}[]
\resizebox{0.45\textwidth}{!}{%
\begin{tabular}{lrl}
\hline
\multicolumn{1}{c}{\textbf{Parameter name}} & \multicolumn{1}{c}{\textbf{Value}} & \multicolumn{1}{c}{\textbf{Units}} \\ \hline
Vibration intensity, $\rho$                 & $5 \times 10^{-4}$                 &                                    \\
Pressure, $p$                               & $40$                               & $MPa$                              \\
Characteristic grain size, $a$              & $350$                              & $\mu m$                            \\
Density, $\rho_g$                           & $1600$                             & $kg/m^3$                           \\
Bulk shear modulus, G                       & $109$                              & $MPa$                              \\
Diffusion constant, $D_0$                   & $1.0$                              &                                    \\
$\hat{\chi}(q)$                             & $0.3$                              &                                    \\
$\tilde{\chi}(\rho | \rho \neq 0)$         & $0.18$                             &                                    \\
Yield stress parameter, $A$                 & $0.30$                             &                                    \\
Yield stress parameter, $B$                 & $3.7 \times 10^{-3}$               &                                    \\
STZ flipping rate, $R_0$                    & $1.00$                             &                                    \\
Vibration constant, $A_0$                   & $0.01$                             &                                    \\
$\epsilon_0$, $\epsilon_z$, $\epsilon_1$    & $1.5$, $0.5$, $0.3$                &                                    \\
Friction constant, $\xi_0$                  & $0.004$                            &                                    \\
Inertial time-scale, $\tau$ & $2.2 \times 10^{-6}$ & \\
Frictional time-scale, $\tau_{fr}$          & $0.013$                            &									\\ \hline                                   
\end{tabular}%
}
\justifying
\caption{Table of simulation parameters. Some simulations with different vibration intensities, $\rho$ were also run. This is indicated before presenting results for the same.}
\label{paramsTable}
\end{table}

A summary of the parameter values used in the following simulations is included in Table \ref{paramsTable}.

\section{Results}
 
\subsection{Steady-state shear stress and compactivity}

\begin{figure}
    \begin{subfigure}[b]{0.45\textwidth}
        \includegraphics[width=1\linewidth]{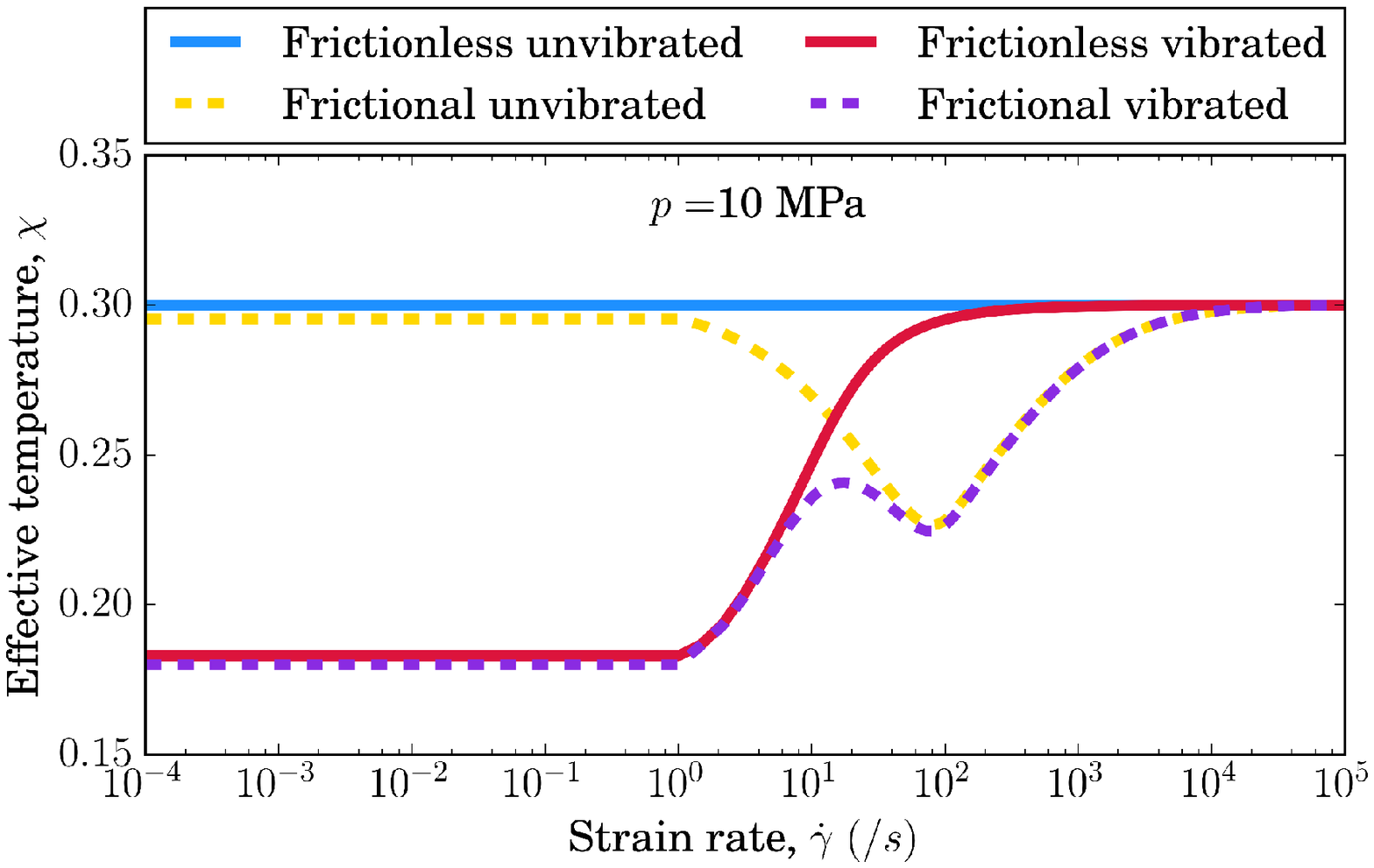}
    \end{subfigure}

    \begin{subfigure}[b]{0.45\textwidth}
        \includegraphics[width=1\linewidth]{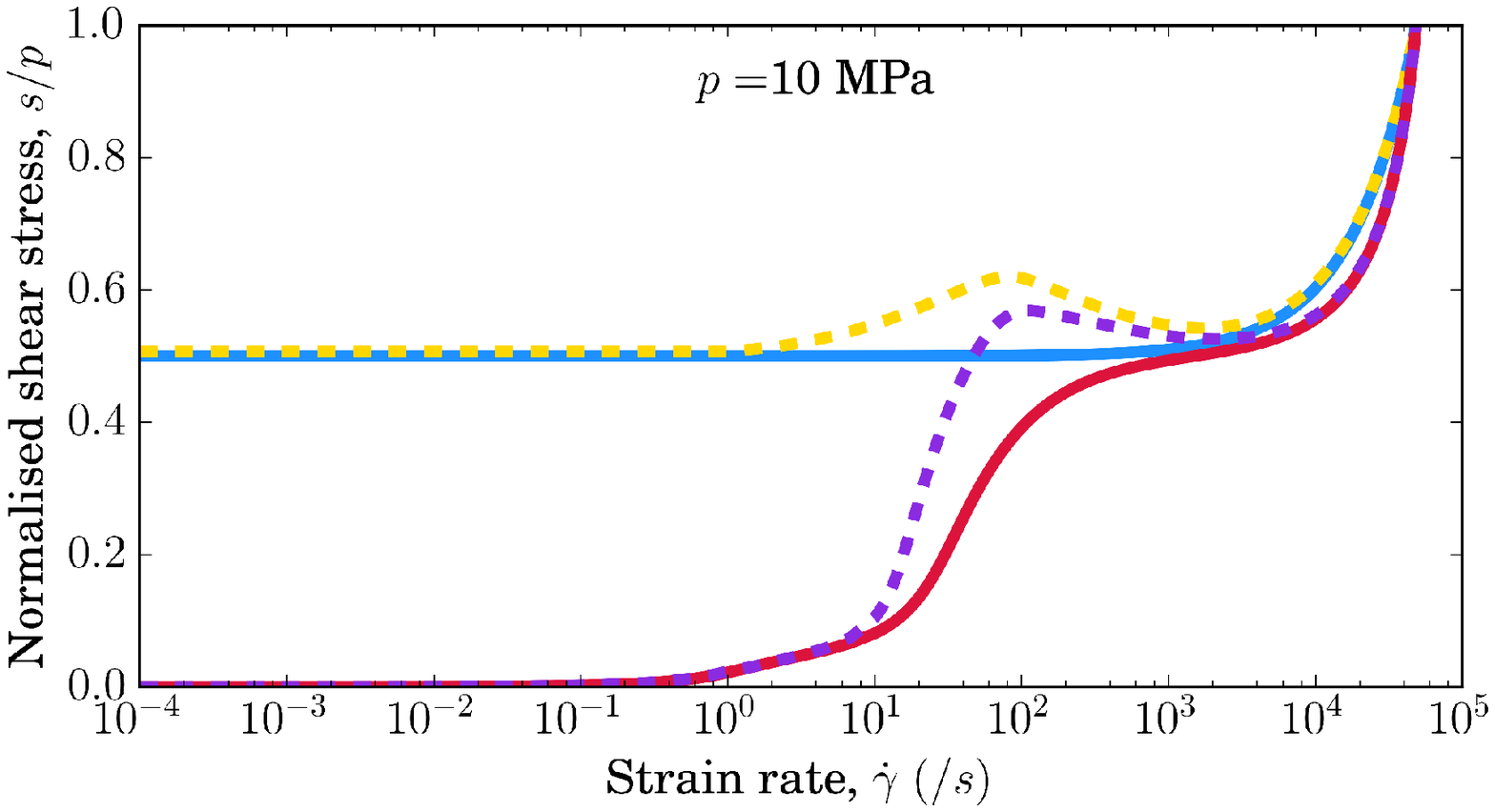}
\end{subfigure}
\caption{Variation of steady-state compactivity and steady-state shear stress values (normalised) with strain rates under different conditions of friction and vibration}
\label{fig:ss_gammadot} 
\end{figure}

We start by solving for the steady-state response of a uniformly sheared layer with no heterogeneity in compactivity, $\chi$. This is done by setting the compactivity evolution rate, $\dot{\chi}$ to zero and the plastic strain rate equal to applied strain rate. We consider $p=10 MPa$ to eliminate the possibility of stick-slip\citep{lieou2015stick}.

In fig. \ref{fig:ss_gammadot}, we clearly see how vibrations fluidizes the system at low strain rates causing compaction of the granular system pushing the compactivity to be closer to $\tilde{\chi}(\rho)$. This effect decreases as the strain rate increases at constant vibration intensity, $\rho$. At low strain rates $\left(\dot{\gamma} < 1/s\right)$, the vibration effects dominate over mechanical loading and we see that the normalised shear stress, goes very close to zero. This implies a regime where the granular network has been completely fluidized. For frictionless grains, vibrations provide a monotonic rate strengthening response. Perhaps, more interesting is the non-monotonicity exhibited by frictional grains. At low strain rates, frictional dissipation allows for compaction of the granular material leading to a slightly lower compactivity. At higher strain rates$\left(\dot{\gamma}\in \left[10^2/s; 4\times10^3/s\right]\right)$, we see strain rate-weakening behaviour which is central to the development of stick-slip instabilities. At large strain rates, all systems converge to a rate strengthening response.

In the upcoming sub-sections, we present our results for dynamic simulations with the formulation described in Eqn.\ref{eqns:all}. We use finite difference methods for spatial derivatives and a predictor-corrector integration scheme for time integration. We chose different simulation scenarios by introducing inter-particle friction and acoustic vibration to the problem one by one and then together. This gives us four conditions viz., no friction and no vibration, only friction, only vibration and both friction and vibration present in the problem. We simulate each condition at 3 different strain rates, $\dot{\gamma}_{pl} = 10/s, 100/s, 1000/s$ signifying different loading rates in the non-monotonic regime.

We divide these results into three sub-sections each evaluating the stress-strain response, the evolution of shear band and the intermittent vibrations respectively.


\begin{figure*}[ht!]
\begin{minipage}{0.30\textwidth}
    \centering
    \begin{subfigure}[b]{\textwidth}
        \includegraphics[width=\textwidth]{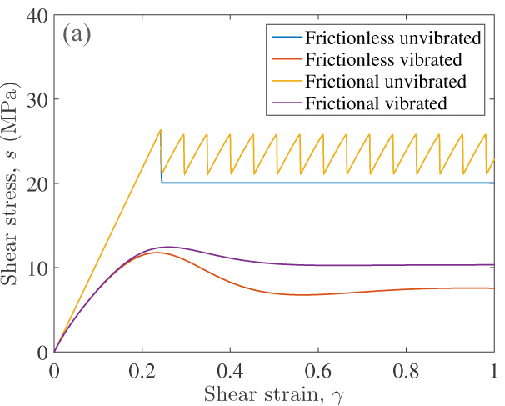}
        \label{10svs}
    \end{subfigure}
\end{minipage}
\begin{minipage}{0.30\textwidth}
    \centering
    \begin{subfigure}[b]{\textwidth}
        \includegraphics[width=\textwidth]{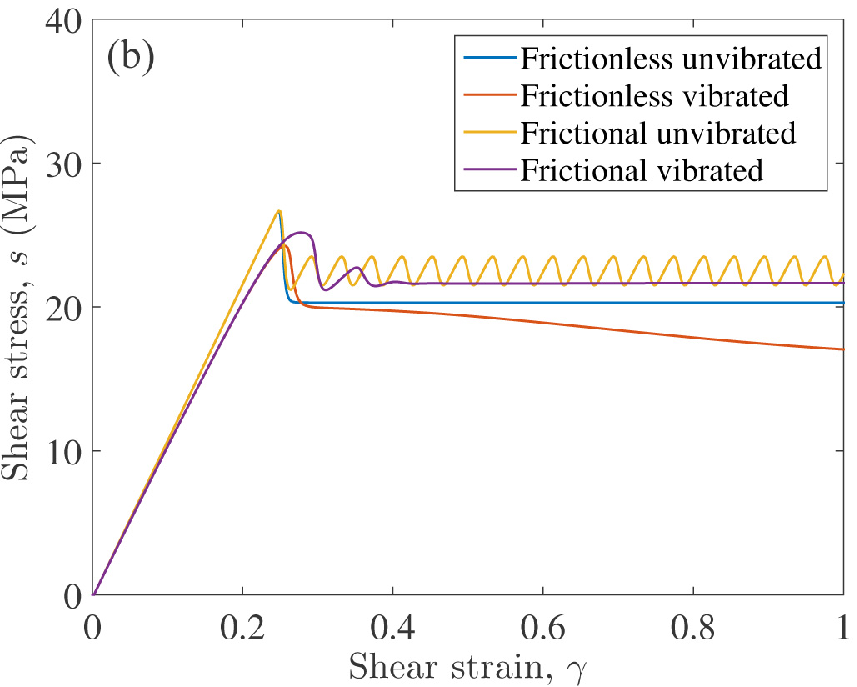}
        \label{100svs}
    \end{subfigure}
\end{minipage}
\begin{minipage}{0.30\textwidth}
    \centering
    \begin{subfigure}[b]{\textwidth}
        \includegraphics[width=\textwidth]{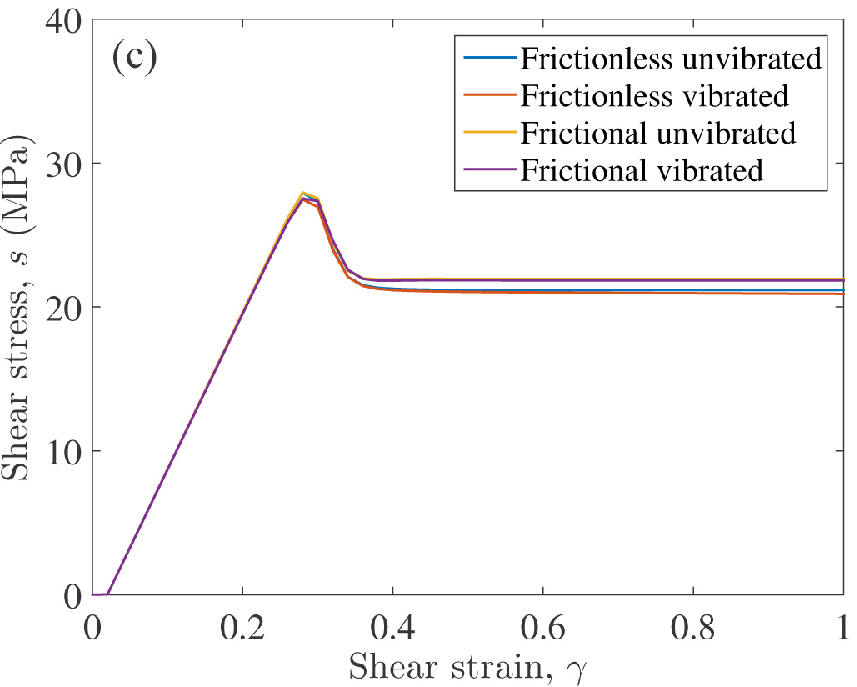}
        \label{1000svs}
    \end{subfigure}
\end{minipage}
\caption{Shear stress variation with respect to shear strain under different simulation conditions (a) $\dot{\gamma} = 10/s$: Stick-slip is observed in the absence of vibrations. Vibrations lowers the peak stress and also the steady-state flow stress (b) $\dot{\gamma} = 100/s$: Stick-slip is observed only in absence of vibration with the response different from (a) in terms of stick-slip cycle times and stress drop magnitudes. (c)  $\dot{\gamma} = 1000/s$: All cases converge to a similar stress-slip response.}
\label{stressVstrain}
\end{figure*}

\begin{figure*}
\begin{minipage}{.31\textwidth}
	\begin{subfigure}{\textwidth}
        \includegraphics[width=\textwidth]{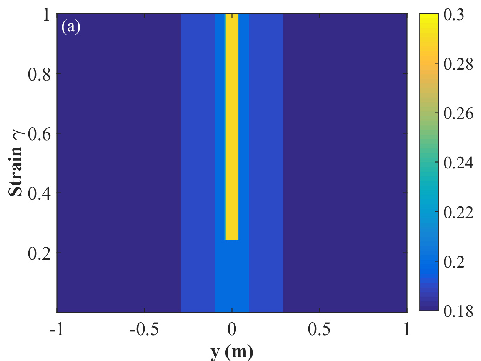}
        \label{fig:chi_40_10_FF}
    \end{subfigure}
    \begin{subfigure}{\textwidth}
        \includegraphics[width=\textwidth]{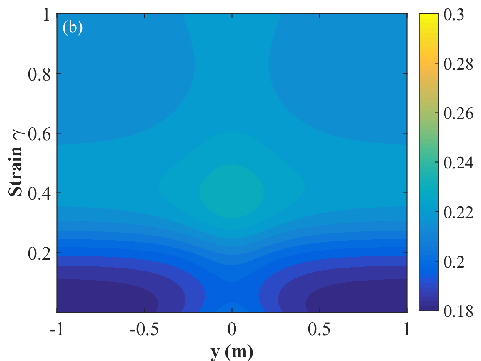}
        \label{fig:chi_40_10_TF}
    \end{subfigure}
    \begin{subfigure}{\textwidth}
        \includegraphics[width=\textwidth]{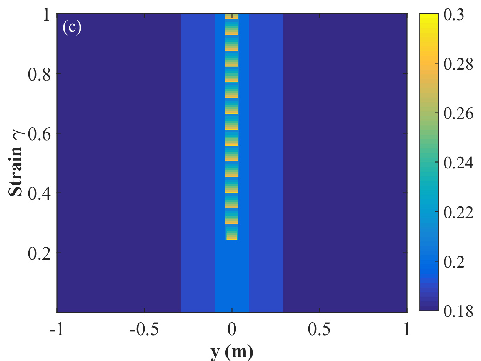}
        \label{fig:chi_40_10_FT}
    \end{subfigure}
    \begin{subfigure}{\textwidth}
        \includegraphics[width=\textwidth]{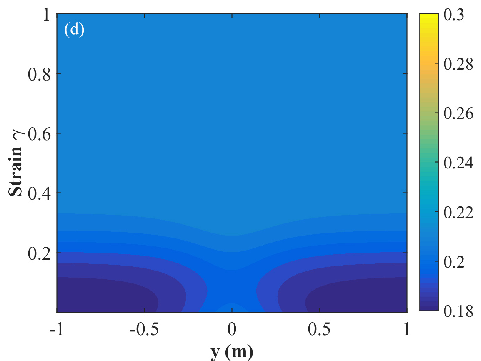}
        \label{fig:chi_40_10_TT}
    \end{subfigure}
\end{minipage}
\begin{minipage}{.31\textwidth}
    \begin{subfigure}{\textwidth}
        \includegraphics[width=\textwidth]{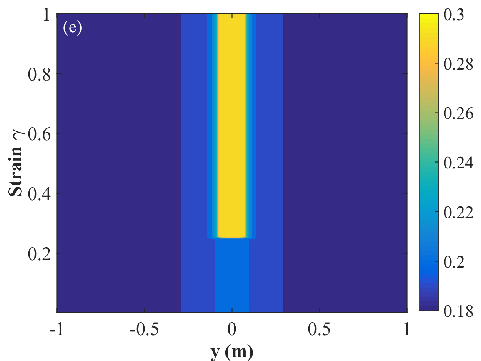}
        \label{fig:chi_40_100_FF}
    \end{subfigure}
    \begin{subfigure}{\textwidth}
        \includegraphics[width=\textwidth]{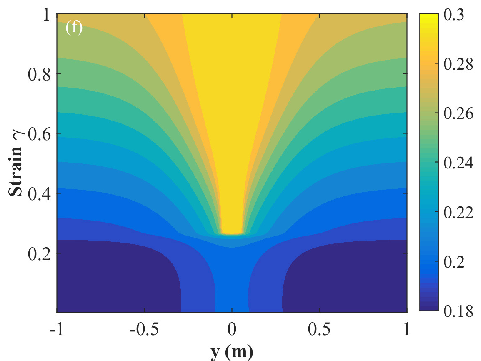}
        \label{fig:chi_40_100_TF}
    \end{subfigure}
    \begin{subfigure}{\textwidth}
        \includegraphics[width=\textwidth]{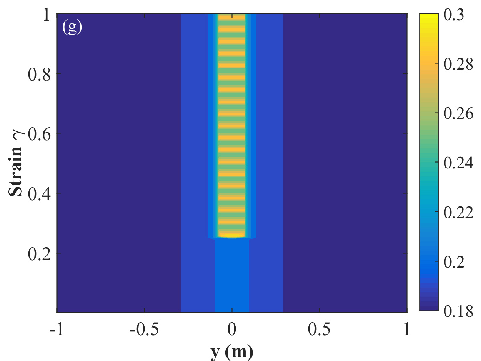}
        \label{fig:chi_40_100_FT}
    \end{subfigure}
    \begin{subfigure}{\textwidth}
        \includegraphics[width=\textwidth]{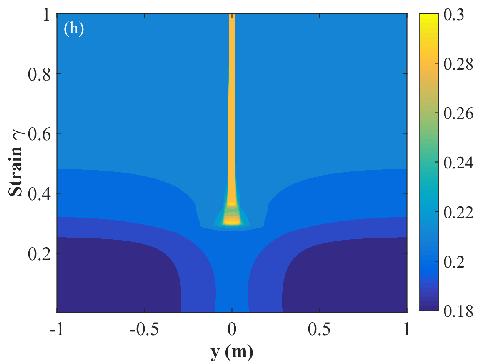}
        \label{fig:chi_40_100_TT}
    \end{subfigure}
\end{minipage}
\begin{minipage}{.31\textwidth}
    \begin{subfigure}{\textwidth}
        \includegraphics[width=\textwidth]{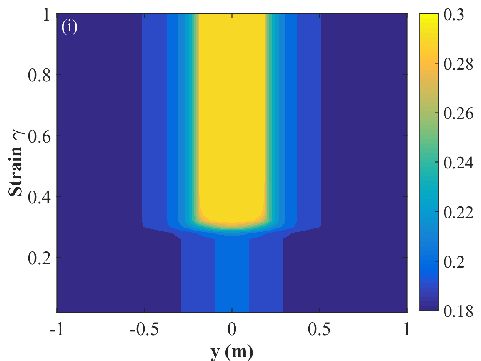}
        \label{fig:chi_40_1000_FF}
    \end{subfigure}
    \begin{subfigure}{\textwidth}
        \includegraphics[width=\textwidth]{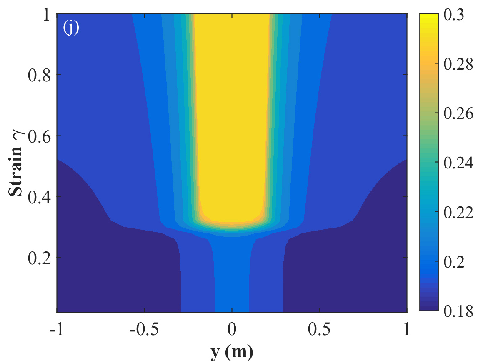} 
        \label{fig:chi_40_1000_TF}
    \end{subfigure}
    \begin{subfigure}{\textwidth}
        \includegraphics[width=\textwidth]{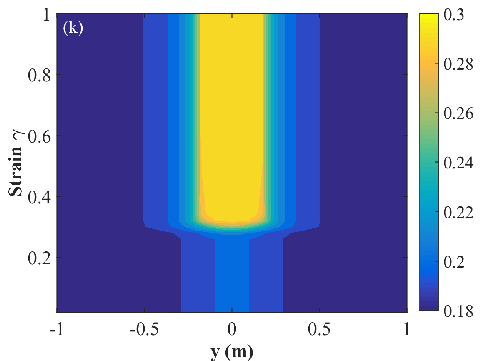}
        \label{fig:chi_40_1000_FT}
    \end{subfigure}
    \begin{subfigure}{\textwidth}
        \includegraphics[width=\textwidth]{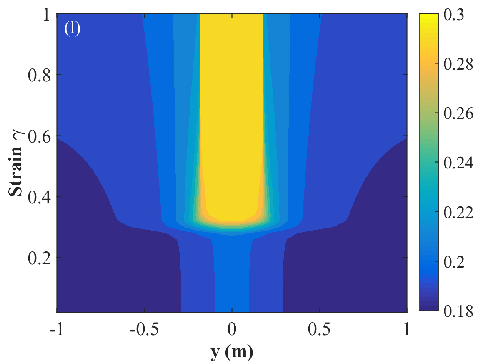}
        \label{fig:chi_40_1000_TT}
    \end{subfigure}
\end{minipage}
\caption{Compactivity, $\chi$ plots: (From L to R) the columns correspond to $\dot{\gamma} = 10/s. \ 100/s, \ 1000/s$ respectively. The $y$-axis in each sub-plot represents increasing strain (time). The $x$-axis represents the spatial co-ordinate. Each row represents a different conditions the simulation was run in: \qquad
\textit{First row}: Frictionless un-vibrated grains,  
\textit{Second row}: Frictionless vibrated grains
\textit{Third row}: Frictional un-vibrated grains
\textit{Fourth row}: Frictional vibrated grains}
Refer to the text for details.
\label{fig:chi_all}
\end{figure*}
\subsection{Stress-strain response}

Figure \ref{stressVstrain} shows the stress-slip response for the sheared granular layer under different conditions and different strain rates.

\textbf{Low strain rates}, $\dot{\gamma} = 10/s$: The shear response varies significantly as different physical processes are considered. For the reference case (frictionless un-vibrated grains), the response is initially linearly elastic. As the shear stress exceeds the yield threshold, plastic strain starts accumulating slowing down the increase in shear stress upto a peak value before it starts to drop. The sharp drop in the stress signifies a rapid increase in the plastic strain rate which leads to strain softening. During the increase in the plastic strain rate post peak yield stress, the compactivity increases until it reaches a steady-state value corresponding to the imposed strain rate(Eqn. \ref{chidot}). In the absence of friction and vibrations, the stress also approaches a steady-state value and does not change with further slip.

The introduction of frictional dissipation significantly alters the response beyond the first peak in stress. For one thing, stick-slip becomes apparent. Previous studies by Lieou \textit{et al.}\citep{lieou2016dynamic,lieou2015stick} on shearing a spatially homogeneous granular layer suggested stick-slip is possible only in the presence of friction. This is still the case even when shear bands form. The stick-slip pattern observed in Fig. \hyperref[10svs]{3a} is periodic. However, the stress drop magnitude is slightly smaller than the stress drop observed in the case of frictionless un-vibrated grains. This may be explained as follows: \\
In the presence of friction, the steady-state compactivity is lower than its corresponding value for the frictionless un-vibrated system (Fig \ref{fig:ss_gammadot}). But the apparent yield stress decreases with increasing compactivity. It follows that at the end of the slip phase, the frictional system reaches a lower compactivity and has a higher yielding threshold than the frictionless system. Hence, the stress drop magnitude is lower.

As for why stick-slip occurs, one should again refer to the Fig. \ref{fig:ss_gammadot}. The curves representing friction show a strain rate-weakening behaviour ($\dot{\gamma}\in \left[10^2/s; 4 \times 10^3\right]$). Since, the shear band is the region where almost all the plastic slip is accommodated, the plastic strain rate within the band is typically much higher than the applied strain rate (Eqn. \ref{sdot}). Hence, even if the applied strain rate, $\dot{\gamma} =10/s$ corresponds to a rate strengthening response, the high strain rate within the shear band pushes the system response to the rate weakening branch leading to emergence of stick slip. This is consistent with previous experimental observations suggesting a transition in sliding stability with the development of shear bands \citep{marone1998laboratory,hadizadeh2015shear}.

Applying external vibrations at an intensity $\rho = 5 \times 10^{-4}$, introduces some notable differences in the stress strain response for both frictional and frictionless grain. First, the linear elastic response is much limited compared to the cases with no vibrations. The stiffness is lower and depends on the accumulated strain signifying the plastic strain has started accumulating sooner. This phenomena may be explained as follows: From Eqn. \ref{meq}, $m_{eq}$ depends explicitly on the vibration intensity, $\rho$. In the case of $\rho \neq 0$, the plastic strain rate computed from Eqn. \ref{gammadot} will no longer exhibit a stability transition at $s \approx s_0$. Rather, plastic strain rate may develop even if $s \leq s_0$; since external vibrations help fluidize the system. Second, the peak stress is reduced in the presence of vibrations. This follows from early accumulation of plastic strain which allows strain softening to start at lower stress and strain values. Third, strain softening is more gradual in the vibrated cases. In the absence of vibrations, the stress drop is sharp and occurs almost instantaneously corresponding to the strain being localized in a relatively narrow shear band post yielding. Vibrations, on the other hand, allow for irreversible transformations in the granular network without requiring any yielding. This leads to the de-localization of strain at low strain rates and broader distribution of plastic flow across the layer width. This in turn leads to a more gradual stress drop.

\textbf{Intermediate strain rates}, $\dot{\gamma} = 100/s$: As the imposed strain rate increases, the differences in the shear response between the different cases diminish. There remain some differences however, which we summarize here. While the response of frictionless un-vibrated layer is almost identical to that at $\dot{\gamma} = 10/s$, the frictional system show stick-slip cycles with smaller stress drop amplitudes and more gradual slip accumulation. The reduced stress drop amplitude is explained using Eqn. \ref{chidot}. In the presence of friction (but no vibrations) $\chi \rightarrow \hat{\chi}(q)\frac{\Gamma}{\Gamma + \xi}$ in order to attain steady-state. As the imposed strain rate increases, the mechanical noise strength, $\Gamma$ increases while $\xi < \xi_0 \ \forall \ \dot{\gamma}_{pl}$. This increases the steady-state compactivity, leading to a reduction in the apparent yield stress. This, in turn, causes the stick phase to develop only up to the reduced yield stress value and hence a decreased amplitude of the stick-slip stress cycle follows. 

As for the comparatively gradual accumulation of deformation during the slip phase, this may be explained by investigating the shear band dynamics. As the imposed strain rate increases, a broader shear band is formed as seen in Fig. \ref{fig:chi_all} to accommodate the plastic slip rate in the system $\left(\dot{\gamma}L = \int_{-L/2}^{L/2} \dot{\gamma}_{pl}dy\right)$ leading to a larger slip-weakening distance. More discussion on the characterization and implications of shear band evolution will follow in the next subsection.

With the application of external vibration, stick-slip cycles diminish in the frictional system and the sliding becomes stable after a few short-lived stick-slip cycles. This corresponds to a brief period of localisation as shown in fig. \hyperref[fig:chi_40_100_TT]{4h}. However, since vibrations allow irreversible plastic strain accumulation without any apparent yield requirement we see that as time passes, the band tends to de-localize and the plastic deformation becomes more distributed across the layer suppressing further stick-slip. For the frictionless system, there is no stick-slip as expected. However, the stress falls steadily post yielding until it gets stabilized and reaches a steady-state at high strains ($\gamma = 7$). A similar trend, but on a shorter strain scale, is observed for the frictionless vibrated case at $\dot{\gamma} = 10/s$.

\textbf{High strain rates}, $\dot{\gamma} = 1000/s$: For the highest strain rate considered in this study, the different cases yield very similar stress-slip response. This may be justified by the fact that in this limit,\textbf{•} the frictional and vibration induced compaction is dominated by shear induced dilation. No stick-slip occurs even in the presence of friction as strain rates within the shear band, despite being higher, are within the rate strengthening limit as shown in Fig. \ref{fig:ss_gammadot}. Vibrations tend to increase compactivity outside the shear band as shown in Fig. \hyperref[fig:chi_40_1000_TF]{4j} and \hyperref[fig:chi_40_1000_TT]{4l} by allowing plastic flow but macroscopic stress values do not change considerably. \\

It should be noted that, at even higher strain rates, it would be more appropriate to use the full Vogel-Fulcher-Tamann (VFT) expression to get the steady-state compactivity as was done in \citep{lieou2014grain, lieou2014shear}. In that case, it is possible that by increasing the noise strength, the system approaches the hydrodynamics limit (infinite compacitivity) and fluidizes. This will alter the response to vibrations and shear beyond what is predicted by the current formulation. Below such fluidization limit, whether the different stress strain curves coincide  will depend on  the accurate characterization of the high strain limit of the frictional term as well as the magnitude of the vibrational time scales compared to the time scale for particle rearrangements. In particular, if the frictional noise (in eqn. \ref{chidot_final}) does not saturate at high strain rates,  the steady state compactivity curves may no longer collapse in that limit.  Furthermore, if the vibrations time scale is smaller than the inverse of the strain rate, the particle rearrangement may be controlled by vibration rather than shear and this will again lead to different responses at high strain rates in the presence and absence of vibrations. This limit of strain rates and vibrations are beyond the scope of this paper.

\subsection{Shear band evolution}

Strain localization may occur in sheared amorphous materials for a variety of reasons. In athermal systems, like the one considered here, shear bands may develop from spatial heterogeneities in the compactivity (effective temperature) distribution. A local increase in the disorder favours higher accumulation of plastic strain rates which in turn lead to further increase in disorder. This positive feedback reinforces the growth of unstable deformation modes leading to the localization of significant portion of plastic strain at locations with highest initial disorder. Non-linear stability analysis by Manning \textit{et al.}\citep{manning2007strain} have shown that if perturbations in effective temperature are high enough, shear bands may develop and persist. However, eventually the disorder diffuses across the layer width and the plastic strain de-localized back into a state of uniform deformation over long time scales.

Here, we analyse strain localization dynamics and the evolution of shear bands in the different systems considered in the previous section. Figure \ref{fig:chi_all} shows the contour plots for the evolution of compactivity across the granular layer with increasing shear strain.
 
\textbf{Low strain rates}, $\dot{\gamma} = 10/s$: Localization occurs only in the absence of vibration. The shear band persists for both the frictional and frictionless systems and its width remains approximately constant throughout the simulation time. For the frictional system, we observe periodic fluctuation in compactivity between high and low values within the banded region. This corresponds to the stick-slip cycles in the stress slip plots of Fig. \hyperref[10svs]{3a}. A slip event corresponds to a sudden increase in compactivity due to sudden shear flow induced dilation. When the sudden dilation is stabilized by frictional compaction we see that the compactivity reduces until the next cycle. It is interesting to note, however, that this stick-slip instability in a spatially heterogeneous system is primarily confined to the shear band region. In a frictionless system, the shear band remains stable.  Applying external vibrations enables the accumulation of plastic strain and increase of compactivity throughout the width of the sheared layer. The strain does not localize and the shear band does not develop.

\begin{figure*}[ht!]
    \begin{subfigure}[b]{0.31\textwidth}
        \includegraphics[width = \textwidth]{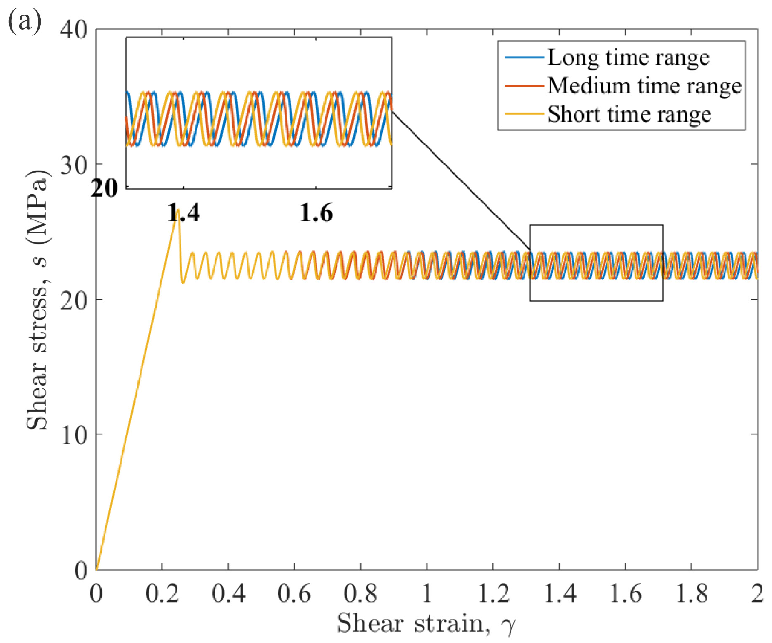}
        \label{low_intVib}
    \end{subfigure}
    \begin{subfigure}[b]{0.31\textwidth}
        \includegraphics[width = \textwidth]{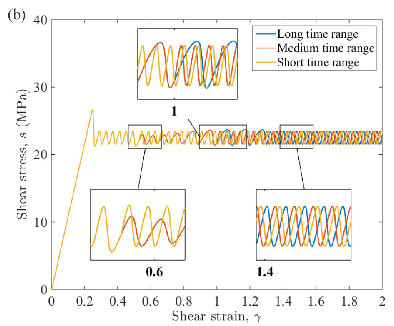}
        \label{med_intVib}
    \end{subfigure}
    \begin{subfigure}[b]{0.31\textwidth}
        \includegraphics[width = \textwidth]{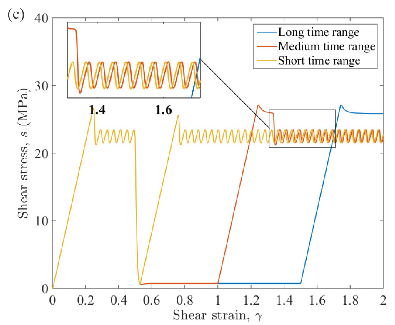}
        \label{high_intVib}
    \end{subfigure}
\caption{Effects of intermittent vibrations on a frictional system sheared at $\dot{\gamma} = 100/s$ (a) \textbf{Low intensity vibrations: }No major changes in stick-slip cycle characteristics are observed for different time exposure ranges.(b) \textbf{Medium intensity vibrations: }Stick-slip cycles are modulated during active vibrations but the original response of the system is recovered after the cessation of vibration. Inserts show relative effects of vibration duration on stress-slip response(c) \textbf{High intensity vibrations: } Complete fluidization of the granular network occurs and stress drops almost completely. If material is exposed for long enough duration, stick-slip is completely eliminated and original response is not recovered after cessation of vibration.}
\label{intVibrations}
\end{figure*}

\textbf{Intermediate strain rates}, $\dot{\gamma} = 100/s$: Similar to the case with lower strain rate, a well localized shear band only forms in the absence of vibration (with the simulation parameters used here). We notice that the shear band thickness increases. Stick-slip is evident in the frictional system with episodes of compaction and dilation within the shear band. For the frictionless systems, the shear band is as stable as before.

The introduction of external vibration enriches the response. For the frictionless system, we observe an initial localization of disorder that gradually spreads out towards a more levelled distribution in which the difference between the maximum and minimum compactivity is just 0.02 at $\gamma = 1.0$. If the simulation is run longer, the vibrations will cause a complete annihilation of this residual inhomogeneity in the disorder. For the frictional system, there is a transient phase of strain localization in which a brief episode of stick-slip occurs. This is reflected in the initial non-smoothness of the localization band. However, as more strain accumulates, the vibrations lead to an increase in compactivity and consequent plastic strain accumulation outside the shear band. This reduces the difference between the compactivity within and outside the shear band and thus the effective degree of localization defined, for example, by the fraction of the plastic flow within the shear band divided by total plastic flow in the entire domain, is reduced compared to the other cases. Hence, while the shear band is apparently narrower in this case, its proportion of plastic flow is lower.

\textbf{High strain rates}, $\dot{\gamma} = 1000/s$: In this limit of high strain rates, all cases exhibit strain localization to different degrees. The shearing is smooth, however, and no stick-slip occurs. The higher imposed strain rate requires a broader shear band to compensate for applied strain rate. The application of external vibration lead to some accumulation of plastic strain outside the shear band as is reflected by increase in compactivity outside the shear bands (Figures \hyperref[fig:chi_40_1000_TF]{4j}, \hyperref[fig:chi_40_1000_TT]{4l}). However, due to the high strain rate, the response is more or less dominated by the mechanical noise strength and the shear band remains well localized throughout the simulation time.

\begin{figure*}
\begin{subfigure}[t]{0.325\textwidth}
    \includegraphics[width=\textwidth]
    {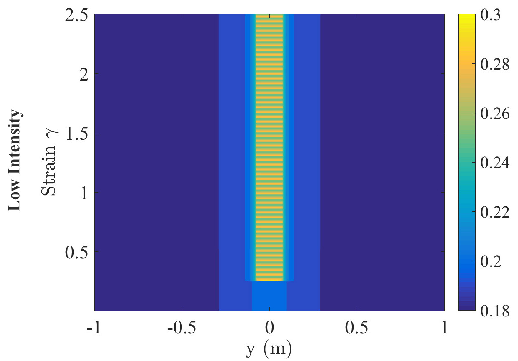}
    \includegraphics[width=\textwidth]
    {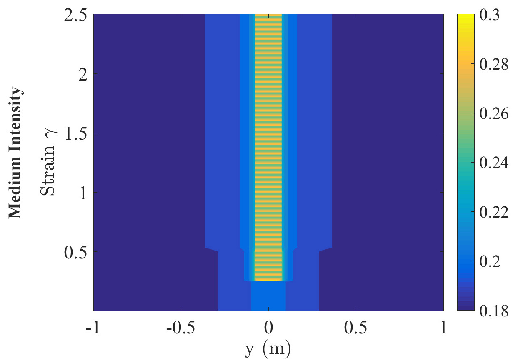}
    \includegraphics[width=\textwidth]
    {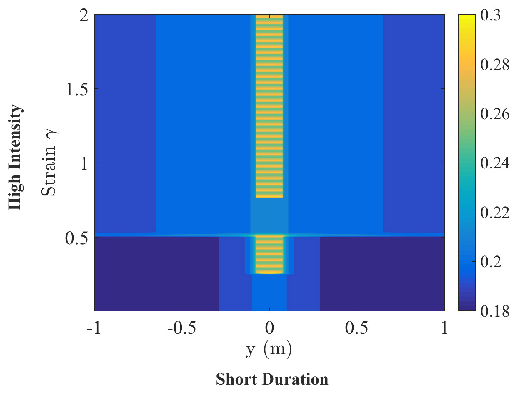}
\end{subfigure}\hfill
\begin{subfigure}[t]{0.30\textwidth}
    \includegraphics[width=\textwidth]  
    {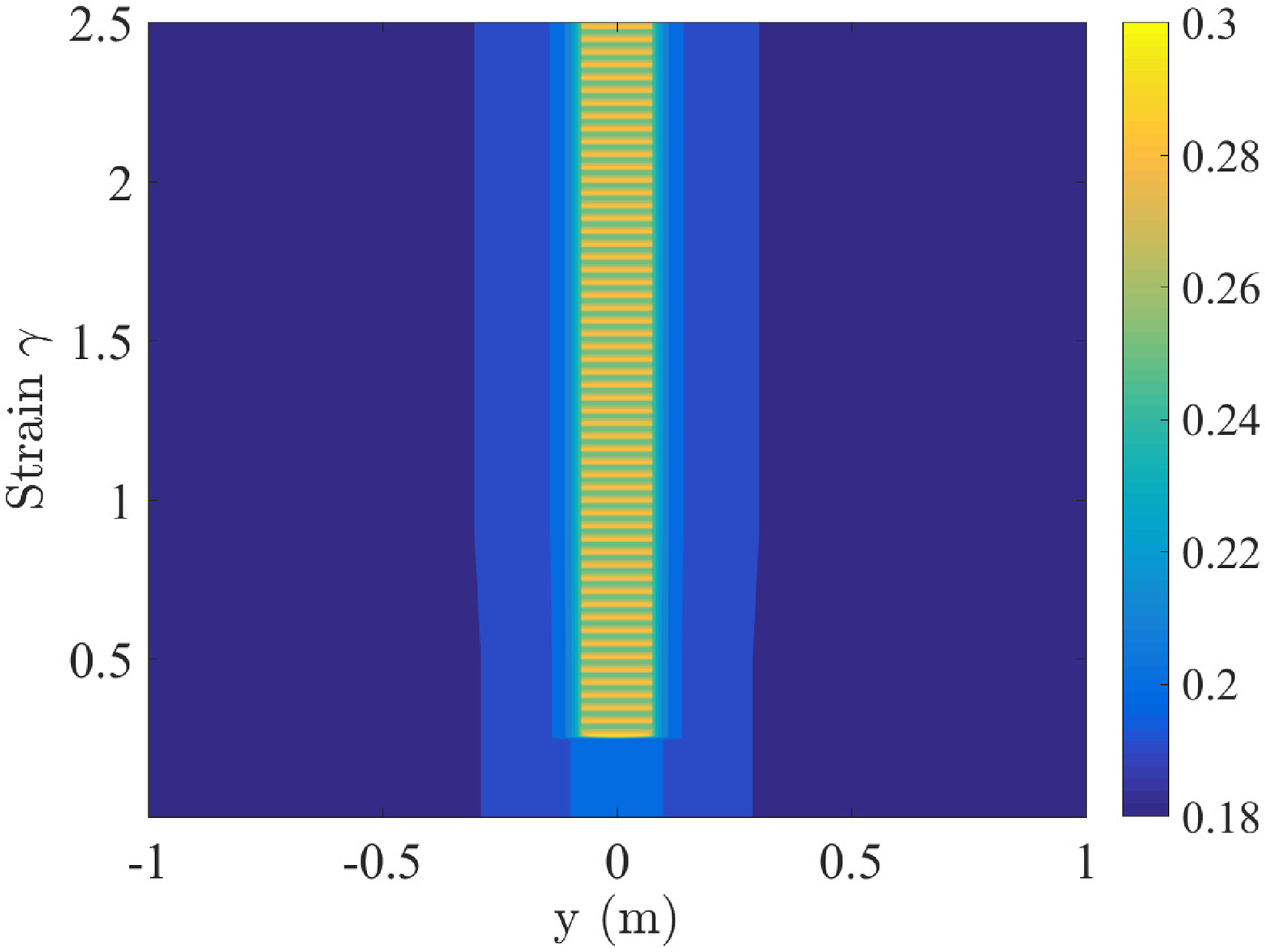}
    \includegraphics[width=\textwidth]
    {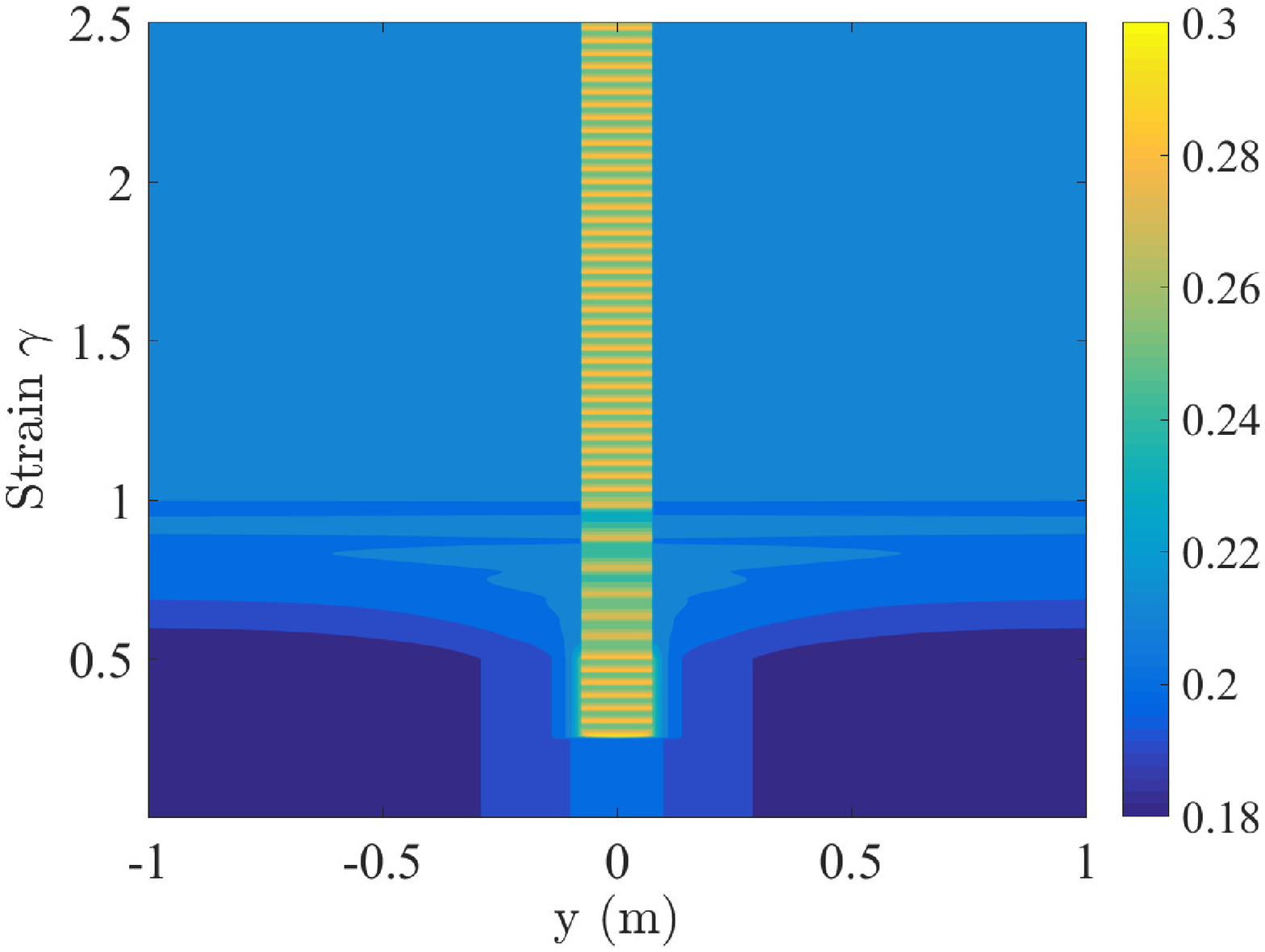}
    \includegraphics[width=\textwidth]
    {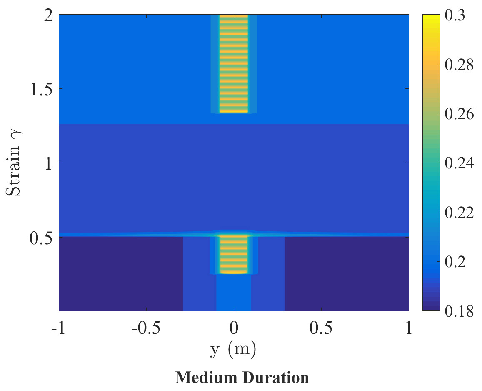}
\end{subfigure}\hfill
\begin{subfigure}[t]{0.30\textwidth}
    \includegraphics[width=\textwidth]  
    {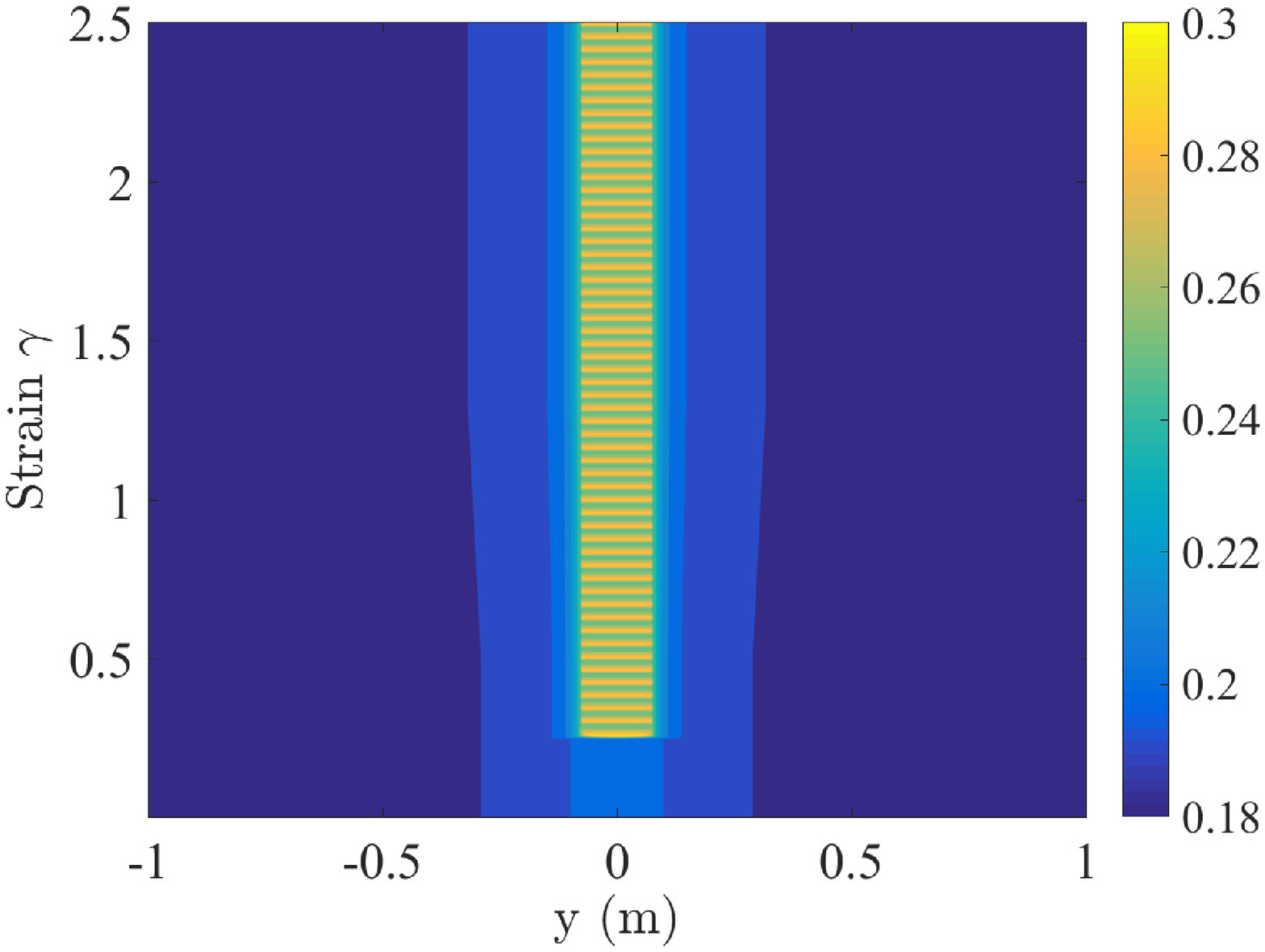}
    \includegraphics[width=\textwidth]
    {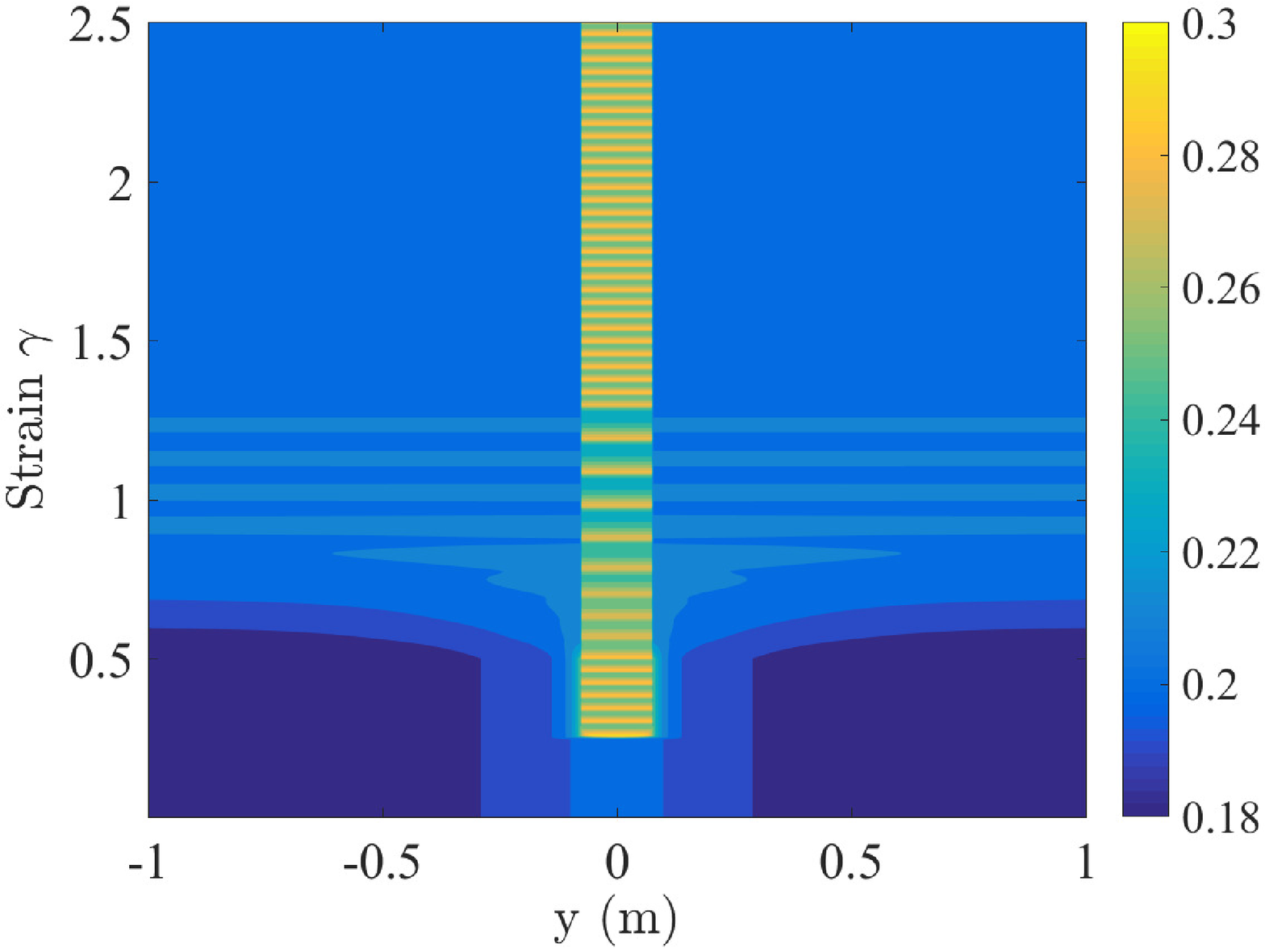}
    \includegraphics[width=\textwidth]
    {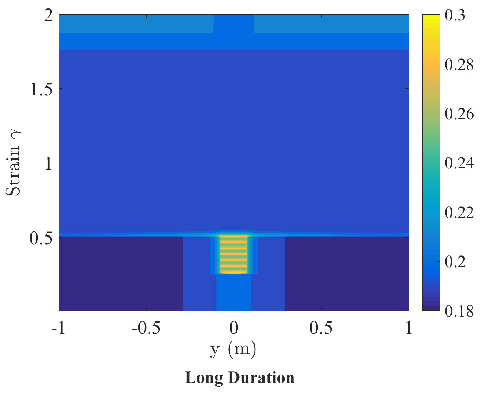}
\end{subfigure}
\caption{{\bf Compactivity distribution plots for a frictional system sheared at $\dot{\gamma} = 100/s$ exposed to different intensities of intermittent vibrations:}(a) Low intensity vibrations cause very minor changes in the compactivity distribution for all three exposure times (b) Medium intensity vibrations change the stick-slip cycle lengths and stress amplitudes. However, stick-slip characteristics are recovered post removal of vibrations (c) A long enough exposure to high intensity vibrations brings about a lasting rheological change in the compactivity distribution eliminating stick-slip. A lower exposure does not completely eliminate the heterogeneity in compactivity and the positive feedback between compactivity and strain rate causes the diminished heterogeneity to grow and form shear bands post removal of vibrations}
\label{iv_chicf}
\end{figure*}

\subsection{Effect of intermittent vibrations}

To test if vibrations can introduce lasting changes in the granular network, a series of simulations with intermittent vibrations were run on the frictional system sheared at strain rate $\dot{\gamma} = 100/s$. Figure \ref{intVibrations} summarizes the results of these numerical experiments. Intermittent vibrations start at strain, $\gamma =0.5$ and end at different strains as evident in Fig. \ref{intVibrations}. The vibrations are applied for three different durations: long, medium and short. Short duration corresponds to vibrations that last less than one stick-slip cycle,  medium duration simulates vibrations that run over a few cycles (5-10) while a long duration vibration is allowed till a new steady-state pattern emerges. The vibrations are not immediately applied at full intensity, $\rho$ instead are ramped linearly from $\rho = 0$ to $\rho = \rho^*$ during a few simulation time-steps. Here, $\rho^*$ is the vibration intensity of the given case.

\textbf{Low intensity vibrations}, $\rho = 5 \times 10^{-6}$: Applying low intensity vibrations after the onset of stick-slip cycles has very little effect on shear response independent of the vibration duration. In particular, the stick-slip pattern does not change during or after the vibration episode. There is a minor difference, however, in the compactivity evolution as shown in Fig. \ref{iv_chicf} for different vibration durations. As the vibration duration increases the plastic strain outside the shear band increases slightly. However, this has no significant implications for the macroscopic stress strain response.

\textbf{Medium Intensity Vibration}, $\rho = 5 \times 10^{-4}$: Applying this level of vibration for a period of just one stick-slip cycle (short duration) has only a minor effect on this particular cycle. As soon as the vibration stops, the stick-slip pattern resumes its pre-vibration trend albeit with some plastic strain accumulated outside the shear band. Increasing the duration of vibration alters the stick slip response. The stress drop per slip event gradually increases (see Fig. \hyperref[med_intVib]{5b}) and the stick-slip duration becomes progressively longer. Plastic strain starts accumulating outside the shear band which increases compactivity outside the band. However, on removal of vibrations, stick-slip continues. A similar trend is observed when the vibration is applied for a long duration. While the stick-slip cycles are modified by the vibration, there are no lasting rheological changes after the vibration is stopped since the shear band characteristics are only slightly affected. 

It may be interesting to compare the results for the intermittent vibrations to those in Fig. \hyperref[100svs]{3b} for which vibrations (of the same intensity), have been applied continuously from the start of shearing. In the continuous vibrations cases, no persistent stick-slip is observed. However, in the intermittent vibrations case, stick-slip characteristics are modified – in terms of its amplitude and frequency- but not diminished. This may be explained as follows: 

Vibrations allow for plastic flow to develop irrespective of the stress value. The amount of plastic flow developed increases gradually with the macroscopic stress and depends on the vibration intensity. If we introduce vibrations from the start, we allow for sufficient plastic flow to be generated across the layer (including outside the region where the shear band may potentially develop). The strain rate may initially localize leading to strain rate values inside the shear band being higher than the imposed strain rate forcing the system into the rate-weakening branch of the stress-slip response and causing stick-slip. However, since plastic flow has been already accumulating outside the shear band due to the sustained vibration, the shear band formed is narrower and has lower degree of localisation (defined as before, fraction of plastic flow rate within the shear band compared to the total plastic flow rate). The change in plastic flow rate in the shear band moves it towards the rate strengthening branch and leads to the disappearance of the transient stick-slip.

Introducing intermittent vibrations for a system that is already in stick-slip has a different effect. During the slip phase, there is a sharp increase in plastic flow rate and compactivity inside the shear band. The local increase in compactivity reduces the yield strength and causes a stress drop in the entire system (since the stress must be uniform everywhere for equilibrium). The reduced macroscopic stress diminishes the vibration induced plastic strain rate outside the shear band. During the stick-phase, the macroscopic stress increases again and plastic flow develops outside the shear band. However, vibration does not have sufficient time to increase the compactivity outside the shear band to a sufficient level that would homogenize the system; the yield threshold is reached within the shear band sooner leading to another episode of slip and instantaneous stress drop. The stick-slip cycle then repeats.

All this suggests that the timing of the application of intermittent vibrations, and what state the system is in apart from the vibration intensity itself, plays a significant role in the system response to vibrations.
 
\textbf{High Intensity Vibration}, $\rho = 5 \times 10^{-2}$: Perhaps this is the case where the most significant changes in the rheology are observed. Due to the high vibration intensity, there is an instantaneous and almost complete stress drop that occurs even if the vibrations are applied for only short period. Furthermore, the shear band almost completely disappears. As a result, after the vibration is stopped, the system does not return instantaneously to its pre-vibrations state. However, it takes some time to reload and reach the yielding threshold again. After reaching this stress peak, the system may still exhibit a different response as shown by the red and blue curves in Fig. \hyperref[high_intVib]{5c}. In particular, no stick-slip is observed following a long time application of transient vibrations. This is attributed to the complete disappearance of the shear band and the attainment of a new steady-state with uniform compactivity. In the case of applying vibration for a medium period, there is a period of stable sliding that follows the cessation of vibrations. However, due to the existence of minor perturbation in the compactivity at the end of the vibration period, this perturbation eventually grows into a new band and the stick-slip pattern is repeated. For the short time vibration, the stick-slip pattern resumes shortly after the elastic reloading phase is over. These results suggest that for a lasting rheological change effect to occur there must be a permanent change in the system state. Here, this change is the disappearance and rebirth of the shear band over different time scales. 

\section{Conclusion}

Understanding deformation and failure in granular materials is a problem of both fundamental importance and practical relevance. This is because many natural phenomena as well as industrial processes are controlled by the physics of granular deformation. Earthquakes, landslides, or pouring, transportation, and mixing in food and pharmaceutical industries are just few examples where granular rheology, especially that which involve the shear response due to local particle rearrangement, is of direct relevance. In many circumstances, the granular deformation is not macroscopically uniform but localizes in shear bands or exhibit spatio-temporal stick-slip complexity. In this paper we presented a numerical model for shearing a granular layer with spatially heterogeneous porosity-like parameter in the presence and absence of external vibrations. We discussed how strain localization and stick-slip instabilities may be possible under these different loading condition.

The consideration of both vibration and shear is motivated by a number of different applications. Recent observations in seismology suggest that dynamic earthquake triggering is a well-documented phenomenon by now\citep{brodsky2014uses,brodsky2005new, wei2015dynamic}. In this case, seismic waves emanating from an earthquake on a given fault may cause premature slip on a distant fault that would not have had moved if the original earthquake did not happen. Seismic waves cause shaking of the fault gouge region, which is already being sheared under the effect of tectonic loading, changing both its rheology and critical stress state, as is suggested by our current model. Another relevant application is control algorithms for transportation of grains in pipelines and on conveyor belts where the sticking of the grains may have detrimental effect on the process. In that case, applying external vibrations should help \textit{un-jam} the granular layer and move it steadily. In this paper, we showed how vibrations and shear interact with each other, affecting dense granular flow by regulating shear band evolution and modifying stick-slip response. The interaction strength depends on the shear rate, vibration intensity, the time of application of vibration and its duration as well as the nature of the grains whether they are frictional or frictionless.

Our primary theoretical tool for this investigation is the shear transformation zone (STZ) theory, a non-equilibrium statistical thermodynamic theory for describing plasticity in amorphous materials. The theory belongs to the family of visco-elastic-plastic models with internal state variables. In the current formulation of the theory, the main state variable- the compactivity, is a measure of volumetric changes in the sheared layer due to the local rearrangements of the particle network. In that sense, a one-to-one correspondence between compactivity evolution and porosity evolution may be established.  

The STZ theory attempts to bridge the macroscale response to the micro/meso-scale disorder. The disorder of the system is captured in its configurational entropy as encapsulated in the compactivity. As noted in \citep{lieou2012nonequilibrium}, compacitivity does behave like an effective temperature in that it is seen as the conjugate to a configurational entropy in the second law variants written for systems modeled with STZs. In recent years, molecular dynamics simulations \citep{manning2011vibrational} have made a lot of progress in identifying these soft-spots and their susceptibility to yielding\citep{patinet2016connecting}. By counting the number of STZs in a given volume, the effective temperature may be directly computed using a Boltzmann distribution in STZ density. This provides a direct microscopic underpinning for the central state variable in the STZ formulation.

The theory was recently modified by \citep{lieou2016dynamic} to incorporate the effects of external vibration on a sheared layer with spatially homogeneous structure. In this paper, we extended this model by allowing the possibility that the spatial distribution of the compactivity or porosity in the granular layer is non-uniform. This enabled the investigation of strain localization and shear band development. Prior studies\citep{segall2006does} developed a porosity-based formulation of gouge plasticity and explored shear banding in that context\citep{marone1998laboratory}. However, to the best of our knowledge this is the first study to consider a statistical thermodynamics-consistent model for that purpose and to incorporate the effects of external vibrations on the shear band dynamics.

In the past few years, there have been some notable attempts to provide an atomistic description for plasticity and yielding in amorphous materials.  For example, \textit{Zacone et al.}\citep{zaccone2014microscopic}  have derived a mean field theory to relate shear induced microscopic configurational nearest neighbor changes directly to homogeneous material properties and stress-strain relations. They provide a simple physical picture in the cage-breaking model: shear causes unequal change in number of neighbors in the compression and extension sectors causing an overall decrease in co-ordination number leading to a decrease in local shear modulus.  The origin of the non-affinity is the force imbalance on groups of particles in the two sectors during shear. Shear forces particles in the extension sector to move away while it forces particles in the contraction sector to move closer in the limit of excluded volume interactions. If one also accounts for viscous stresses using a viscoelastic (Zener) solid model, one can account for the stress overshoot existence and amplitude due to the competition between non-affine shear induced cage-breakup and the build-up of viscous stresses. 

While the STZ theory does not explicitly account for the detailed force balance in individual STZs, it connects to the microscopic picture outlined above by accounting for entropy evolution and energy balance (see \hyperref[AppendixA]{Appendix}). Furthermore, the inelastic strain rate is consistently derived based on a simple kinematic picture of STZ transitions. With this thermodynamics-based formulation it was possible to account for homogeneous and inhomogeneous deformation in a variety of amorphous systems in both thermal and athermal systems \citep{falk1998dynamics, langer2003dynamics, langer2004dynamics, langer2008shear}. Nonetheless, more work is required to combine recent advances in atomistic and molecular  thories of flow in amorphous materials with the statistical thermodynamic framework of the STZ theory to provide a truly multiscale framework.

Our results suggest that strain localization occurs in both frictional and frictionless systems in the absence of vibration. The application of external vibrations may prevent the formation of the shear band altogether at low strain rates. As the strain rates increases, the competition between shearing and vibration may allow shear band development at low strain but causes de-localization of slip at larger strains.  In the limit of high strain rates, the strain localizes due to the dominance of mechanical noise which the external vibration may not overcome (for the vibration intensity values considered here). These results may provide guidelines in controlling physical systems incorporating friction or sliding on granular/powder layers where strain localization is not desirable by changing the level of applied external vibrations.

Our results further suggest that stick-slip occurs only in systems with frictional grains. This is consistent with the findings of Lieou \textit{et al.} \citep{lieou2015stick,lieou2016dynamic}. However, due to the strain localization in our simulations the stick-slip is only confined to the banded domain. A signature of this is the episodes of periodic dilation and compaction seen in the compactivity distribution within the shear band in Figs. (\hyperref[fig:chi_40_100_FT]{4g}, \hyperref[fig:chi_40_10_FT]{4c}). Recent experiments on bulk metallic glass by authors of \citep{maass2015long} reported that cavitation and roughness are observed within the shear band upon investigating post fracture surfaces. Our results provide a framework for explaining these observations in terms of local compaction and dilation in the presence of rate dependent dissipative processes (The frictional noise term used here is one possible candidate for these internal dissipative phenomena).

Numerous laboratory experiments suggested that strain localization is associated with a transition into rate weakening response as measured by the (a-b) rate parameters in velocity stepping experiments \citep{marone1998laboratory,hadizadeh2015shear}. Several hypothesis have been put forward to explain this transition including grain fragmentation and rotations facilitating grain alignment in the direction of shear. Our model suggests that this possible transition associated with strain localization may not be universal but will happen in a given range of strain rates due to the auto-acoustic friction induced compaction (which may be very well due to grain collision, rotation and alignment). In particular, auto-acoustic compaction leads to non-monotonic rheology as suggested by recent experimental results of \citep{elst2012auto}. The local increase in strain rate within the shear band may shift the system response from a rate strengthening branch to a rate weakening branch even if the imposed strain rate entails a rate strengthening behaviour. For these very same reasons, it is possible that strain localization occurs and the stability of sliding does not change (i.e. remains rate strengthening). We have observed this to be the case for high strain rates where the response is already on the rate strengthening branch at both the low (imposed) and high (local) strain rates. This points to the role of local processes and heterogeneities in controlling macroscopic response and overall sliding stability.

The stick-slip cycles may change due to the application of external vibrations. If vibrations are applied during the early elastic loading phase of the granular layer the stick-slip may completely diminish resulting in long-term stable sliding. However, if external vibrations are applied after the development of the stick-slip cycle, vibrations may alter the stick-slip response. The effect depends on the duration and intensity of external vibration. Short duration low amplitude vibrations has nearly no effect. However, long duration large intensity vibration may abolish the stick-slip cycles, lead to large stress drops, and lead to stable sliding even after the cessation of vibration. This may be considered as another example of vibration-induced lasting rheological changes similar to those that have been observed in recent discrete element models \citep{johnson2013acoustic} and laboratory experiments  \citep{johnson2008effects,johnson2016dynamically}.  

In seismology, it has been observed that elastic waves may, in some cases, cause early triggering of slip and amplification of seismic activity  \citep{brodsky2000new,manga2006seismic}. In some other cases, it has been reported that vibrations may lead to quiescence and a period of no seismicity. Our results suggest that vibrations may amplify or diminish stick-slip cycles within the shear band depending on the state of the system at the time of application of the vibrations and the properties of the vibrations. For example, applying medium intensity vibration for an extended period of time may amplify the stress drop during the slip phase and prolong the slip duration. On the other hand, large intensity vibration applied for an extended period of time may lead to a quiescent period where steady sliding prevails following the cessation of vibrations. A new cycle of stick-slip may follow this quiescent period depending on the duration of the vibration phase. The longer the vibration period, the less probable that stick-slip will re-emerge.

Lasting rheological changes in this study is primarily due to permanent changes that the vibrations introduce into the structure of the shear band. The broadening or disappearance of the shear band as a result of external vibrations changes the state of the system at the end of the vibration period. It takes time for the strain to re-localize again, if at all possible. This leads to changes in the post-vibration period rheology. However, other candidates for inducing lasting rheological changes are also possible. For example, changes in elastic modulus due to vibration induced compaction is a possibility. Another possibility is viscous relaxation and ageing due to creep at inter-particle contacts.  This may be incorporated in our model through a two scale model similar to the one used by Elbanna and Carlson, 2014 \citep{elbanna2014two} in which physics of sliding at contact asperities informs the STZ model. A third possibility is that physical systems of interest (e.g. faults) extend spatially and non-uniform interaction between the different components of such systems, that are subjected to inhomogeneous vibration intensity, may be responsible for different rheological response after the cessation of vibrations.

Future work may focus on the extension of the current 1D model into 2D model for gouge layers that will enable more complex evolution of the shear band and the incorporation of complex spatial heterogeneities. Another possible extension is the incorporation of the STZ model with vibration dependent rheology into a  simplified dynamic rupture model such as the Burridge-Knopoff system for which the long term failure statistics may be explored and statistics of event sizes, stress drops, and inter-event times may be investigated.

{\bf Acknowledgement: } This work has been supported by the National Science Foundation (CMMI-1435920)  and (EAR-1345108).

\appendix*
\section{Thermodynamic origins of STZ theory} \label{AppendixA}

In this section, we attempt to provide a concise summary of the STZ thermodynamic formulation. Much of this section follows from \citep{lieou2012nonequilibrium}.

The STZ theory assumes a (weak) coupling of two subsystems: a \textit{slow} configurational subsystem that incorporates different arrangements of the individual particles in the bulk material and a \textit{fast} kinetic subsystem that includes the vibrational degrees of freedom. Suppose that such a system is in contact with a thermal reservoir that is at a constant temperature $\theta$. We can then write the first law of thermodynamics for this system under a pressure, $p$ and a shear stress, $s$ as \citep{lieou2012nonequilibrium}:
\begin{align}
\dot{U}_T &= Vs\dot{\gamma} - p\dot{V}(S_C,\zeta_i)\\
&= Vs\dot{\gamma} - pX\dot{S}_C - p \left(\dfrac{\partial V}{\partial \zeta_i}\right)_{S_C} \dot{\zeta}_i	
\end{align}
Here, $\dot{\gamma}$ is the strain rate, $X$ is defined in \ref{unnorm_chidef}, $\zeta_i$ are the internal state variables representing the configuration states and the summation convention is implied on $\zeta_i$'s. Let $S_T$ denote the entropy of kinetic subsystem along with that of the reservoir. Then, writing $\dot{U}_T = \theta \dot{S}_T$, we have:
\begin{equation} \label{scdot}
pX\dot{S}_C = Vs\dot{\gamma} - p \left(\dfrac{\partial V}{\partial \zeta_i}\right)_{S_C} \dot{\zeta}_i	 - \theta \dot{S}_T
\end{equation}
Invoking the second law, we know that $\dot{S} = \dot{S}_C + \dot{S}_T \geq 0$. Therefore, using eqn. \ref{scdot}, one can show:
\begin{subequations}
\begin{equation} \label{dissineq}
Vs\dot{\gamma} - p \left(\dfrac{\partial V}{\partial \zeta_i}\right)_{S_C} \dot{\zeta}_i \geq 0
\end{equation}
\begin{equation} \label{qineq}
(pX- \theta)\dot{S}_T \geq 0
\end{equation}
\end{subequations}

Equation \ref{dissineq} signifies that the dissipation be non-negative. It is the difference between the rate of mechanical work done on the configurational subsystem and the rate of energy stored in the internal degrees of freedom of the same subsystem. The second inequality (eq. \ref{qineq}) implies that both $\dot{S}_T$ and $pX-\theta$ carry the same sign, so that
\begin{equation} \label{thetast}
\theta \dot{S}_T  = -\kappa \left( 1 - \dfrac{pX}{\theta}\right)
\end{equation}
where, $\kappa$ is a non-negative thermal transport coefficient and the right hand side denotes the flux from the configurational subsystem to the reservoir.

For STZs, we assume a two-state defect '+' and '-'. The internal state variables are as defined in equation \ref{micro}. If $N$ is the total number of STZs and $v_z$ the excess volume per STZ, then:
\begin{equation}
V = N\Lambda v_z + V_1(S_C - S_z(\Lambda , m)) 
\end{equation}
where $S_z$ denotes the entropy associated with the STZs. Then, from \citep{bouchbinder2009nonequilibrium}:
\begin{equation} \label{voleq}
S_z(\Lambda, m) = NS_0(\Lambda) + N\lambda\psi(m)
\end{equation}
where
\begin{equation} \label{entropy0}
S_0(\Lambda) = -\Lambda ln \Lambda  + \Lambda  
\end{equation}

\begin{equation}\label{psim}
\psi(m) = ln 2 - \frac{1}{2}(1+m)ln(1+m) -\frac{1}{2}(1-m)ln(1-m)
\end{equation}

Using eqns. \ref{thetast}, \ref{voleq}, \ref{entropy0}, \ref{psim} in eqns \ref{micro_all_rate} to write the dissipation $\mathcal{W}$ in terms of the state variables would then be:

\begin{align} \label{full_dissipation_eqn}
\dfrac{\tau \mathcal{W}}{v_z N} = & -\tilde{\Gamma}p\chi\Lambda m \dfrac{d\psi}{dm}\notag\\ & + 2\Lambda C(\mu, \chi)(T(\mu, \chi) - m)\left(\dfrac{v_0}{v_z} s + p\chi\dfrac{d\psi}{dm}\right)\notag\\
& - p\tilde{\Gamma}(\Lambda^{eq} - \Lambda)\left(1 + \chi \left(ln \Lambda - \psi(m) + m \dfrac{d\psi}{dm}\right)\right)
\end{align}

Since, $\mathcal{W} \geq 0$, all the three terms should be individually non-negative. The first term is satisfied by definition of $\psi(m)$ as:
\begin{equation}
\dfrac{d\psi}{dm} = -\dfrac{1}{2}ln \left(\dfrac{1+m}{1-m}\right) = -tanh^{-1}(m)
\end{equation}
which implies that $-m\dfrac{d\psi}{dm}$ is automatically negative. For the non-negativity of second term we need:

\begin{equation}
\mathcal{T} = tanh \left(\dfrac{v_0 s}{v_z p\chi}\right) = tanh \left(\dfrac{\epsilon_0 \mu}{\epsilon_z \chi}\right)
\end{equation}

This is a notable finding, since it constrains the rate factors from a thermodynamics standpoint. Recalling the definition of $\mathcal{T}$ and noting the symmetry of rate factors, we assume $R(\mu, \chi)\approx exp\left(\frac{\epsilon_0 \mu}{\epsilon_z\chi}\right)$. Thus, higher stresses activate more transitions, while higher configurational disorder is reflective of a very noisy system; thereby reducing the transition rate.

Lastly, the third term of eqn. \ref{full_dissipation_eqn} is modelled as $-\dfrac{\partial F(\Lambda, m)}{\partial \Lambda} (\Lambda^{eq} - \Lambda) \geq 0$, assuming $F(\Lambda, m)$ represents the free energy of the granular subsystem. Using the non-negativity condition for the third term, we get:
\begin{equation}
\Lambda^{eq} = exp\left(-\dfrac{1}{\chi} + \psi(m) - m\dfrac{d\psi}{dm}\right) \approx 2exp\left(-\frac{1}{\chi}\right)
\end{equation}
Thus, the equilibrium density of STZs is given by a Boltzmann-like factor as shown above.

\bibliography{Kothari2016}
\end{document}